\documentclass[a4paper,fleqn,usenatbib]{mnras}
\usepackage{latexsym}
\usepackage{amsmath}
\usepackage{amssymb}
\usepackage{fnpos}
\usepackage{hyperref}
\usepackage{tabularx}
\usepackage{xspace}
\usepackage{enumerate}
\usepackage{scalerel}
\usepackage{graphicx}
\usepackage{url}
\usepackage{caption}
\usepackage{gensymb}
\usepackage{longtable}
\usepackage{lscape}
\usepackage[normalem]{ulem} 
\usepackage[dvipsnames]{xcolor} 
\usepackage{wasysym} 
\definecolor{darkgreen}{rgb}{0.0,0.55,0.0}
\definecolor{darkblue}{rgb}{0.0,0.0,0.5}

\newcommand{\eal}[2]{\ifmmode{\mathrm{#1\,#2}}\else{#1\textsc{$\,$\lowercase{#2}}}\fi\xspace}
\newcommand{\feal}[2]{\ifmmode{\mathrm{#1\,#2}}\else{[#1\textsc{$\,$\lowercase{#2}}]}\fi\xspace}
\newcommand{\hfeal}[2]{\ifmmode{\mathrm{#1\,#2}}\else{#1\textsc{$\,$\lowercase{#2}}]}\fi\xspace}

\newcommand{\orcid}[1]{$^{\rm \href{https://orcid.org/#1}{\includegraphics[height=0.6em]{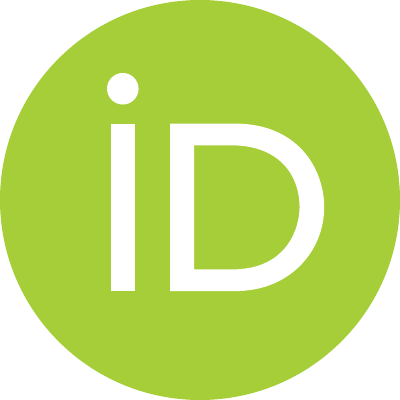}}}$}

\title[The Fe II and He/N spectral phases of novae]{Revisiting the classics: On the evolutionary origin of the ``Fe II'' and ``He/N'' spectral classes of novae}
\author[Aydi et al.]{\parbox{\textwidth}{E.~Aydi\orcid{0000-0001-8525-3442}$^{1}$\thanks{Elias Aydi - NFHP Hubble Fellow; E-mail: aydielia@msu.edu}, L.~Chomiuk\orcid{0000-0002-8400-3705}$^{1}$, J.~Strader$^{1}$, K.~V.~Sokolovsky\orcid{0000-0001-5991-6863}$^{2, 3}$, R.~E.~Williams$^{4, 5}$, D.~A.~H.~Buckley$^{6, 7}$, A.~Ederoclite\orcid{0000-0003-3656-5524}$^{8}$, L.~Izzo\orcid{0000-0001-9695-8472}$^{9}$, R.~Kyer\orcid{0009-0006-2411-5162}$^{1}$, J.~D.~Linford\orcid{0000-0002-3873-5497}$^{10}$, A.~Kniazev\orcid{0000-0001-8646-0419}$^{6, 11, 3, 12}$, B.~D.~Metzger\orcid{0000-0002-4670-7509}$^{13, 14}$, J.~Miko{\l}ajewska\orcid{0000-0003-3457-0020}$^{15}$, P.~Molaro$^{16, 17}$, I.~Molina$^{1}$, K.~Mukai$^{18, 19}$, U.~Munari$^{20}$, M.~Orio$^{21, 22}$, T.~Panurach\orcid{0000-0001-8424-2848}$^{1, 23}$, B.~J.~Shappee\orcid{0000-0003-4631-1149}$^{24}$, K.~J.~Shen\orcid{0000-0002-9632-6106}$^{25}$, J.~L.~Sokoloski$^{26}$, R.~Urquhart$^{1}$, and F.~M.~Walter\orcid{0000-0001-7796-1756}$^{27}$}
\vspace{0.4cm}\\
\parbox{\textwidth}{
$^{1}$Center for Data Intensive and Time Domain Astronomy, Department of Physics and Astronomy, Michigan State University, East Lansing, MI 48824, USA\\
$^{2}$Department of Astronomy, University of Illinois at Urbana-Champaign, 1002 W. Green Street, Urbana, IL 61801, USA\\
$^{3}$Sternberg Astronomical Institute, Moscow State University, Universitetskii~pr.~13, 119992~Moscow, Russia\\
$^{4}$Department of Astronomy \& Astrophysics, University of California, Santa Cruz, 1156 High Street, Santa Cruz, CA 95064, USA\\
$^{5}$Space Telescope Science Institute, 3700 San Martin Drive, Baltimore, MD 21218, USA\\
$^{6}$South African Astronomical Observatory, P.O.\ Box 9, 7935 Observatory, South Africa\\
$^{7}$Department of Astronomy, University of Cape Town, Private Bag X3, Rondebosch 7701, South Africa\\
$^{8}$Centro de Estudios de F\'isica del Cosmos de Arag\'on, Plaza San Juan 1, 44001, Teruel, Spain\\
$^{9}$UK Astronomy Technology Centre, Royal Observatory, Blackford Hill, Edinburgh, EH9 3HJ, UK\\
$^{9}$DARK, Niels Bohr Institute, University of Copenhagen, Jagtvej 128, 2200 Copenhagen {\O}, Denmark \\
$^{10}$National Radio Astronomy Observatory, P.O.\ Box O, Socorro, NM 87801, USA\\
$^{11}$Southern African Large Telescope Foundation, PO Box 9, Observatory 7935, South Africa\\
$^{12}$Special Astrophysical Observatory, Nizhnij Arkhyz, Karachai-Circassia, 369167, Russia\\
$^{13}$Department of Physics and Columbia Astrophysics Laboratory, Columbia University, New York, NY 10027, USA\\
$^{14}$Center for Computational Astrophysics, Flatiron Institute, 162 5th Ave, New York, NY 10010, USA\\
$^{15}$Nicolaus Copernicus Astronomical Center, Polish Academy of Sciences, Bartycka 18, PL 00-716 Warsaw, Poland
$^{16}$INAF-Osservatorio Astronomico di Trieste, Via G.B. Tiepolo 11, I-34143 Trieste, Italy\\
$^{17}$Institute of Fundamental Physics of the Universe, Via Beirut 2, Miramare, Trieste, Italy\\
$^{18}$CRESST and X-ray Astrophysics Laboratory, NASA/GSFC, Greenbelt, MD 20771, USA\\
$^{19}$Department of Physics, University of Maryland, Baltimore County, 1000 Hilltop Circle, Baltimore, MD 21250, USA\\
$^{20}$INAF Astronomical Observatory of Padova, 36012 Asiago (VI), Italy\\
$^{21}$INAF--Osservatorio di Padova, vicolo dell'Osservatorio 5, I-35122 Padova, Italy\\
$^{22}$Department of Astronomy, University of Wisconsin, 475 N.\ Charter St., Madison, WI 53704, USA\\
$^{23}$Center for Materials Research, Department of Physics, Norfolk State University, Norfolk VA 23504, USA\\
$^{24}$Institute for Astronomy, University of Hawai'i, 2680 Woodlawn Drive, Honolulu, HI 96822, USA\\
$^{25}$Department of Astronomy and Theoretical Astrophysics Center, University of California, Berkeley, CA 94720, USA\\
$^{26}$Columbia Astrophysics Laboratory and Department of Physics, Columbia University, New York, NY 10027, USA\\
$^{27}$Department of Physics \& Astronomy, Stony Brook University, Stony Brook, NY 11794-3800 USA\\
}}

\pubyear{2023}

\begin{document}
\label{firstpage}
\pagerange{\pageref{firstpage}--\pageref{lastpage}}
\maketitle

\begin{abstract}
The optical spectra of novae are characterized by emission lines from the hydrogen Balmer series and either Fe~II or He/N, leading to their traditional classiﬁcation into two spectral classes: “Fe~II” and “He/N”. For decades, the origins of these spectral features were discussed in the literature in the contexts of different bodies of gas or changes in the opacity of the ejecta, particularly associated with studies by R.~E.~Williams and S.~N.~Shore. Here, we revisit these major studies with dedicated, modern data sets, covering the evolution of several novae from early rise to peak all the way to the nebular phase. Our data confirm previous suggestions in the literature that the “Fe II” and “He/N” spectral classes are phases in the spectroscopic evolution of novae driven primarily by changes in the opacity, ionization, and density of the ejecta, and most if not all novae go through at least three spectroscopic phases as their eruptions evolve: an early He/N (phase 1; observed during the early rise to visible peak and characterized by P Cygni lines of He~I and N~II/III), then an Fe~II (phase 2; observed near visible peak and characterized by P Cygni lines of Fe~II and O~I), and then a later He/N (phase 3; observed during the decline and characterized by emission lines of He~I/II, N~II/III), before entering the nebular phase. This spectral evolution seems to be ubiquitous across novae, regardless of their speed class; however the duration of each of these phases differs based on the speed class of the nova.
\end{abstract}


\section{Introduction}

A classical nova is a thermonuclear eruption occurring in the hydrogen-rich shell formed on the surface of a white dwarf star accreting material from a nearby stellar companion (for reviews see, e.g., \citealt{Gallagher_Starrfield_1976,Bode_etal_2008,Della_Valle_Izzo_2020,Chomiuk_etal_2020}). 
The thermonuclear eruption 
produces an optical transient, where the system brightens by 8 up to 15 magnitudes, reaching naked-eye brightness in some cases. 

For decades, nova eruptions were monitored using optical spectroscopy, where emission and absorption lines of a wide diversity of species are observed, characterized by velocities ranging from a few hundreds up to a few thousands km\,s$^{-1}$ 
(e.g., \citealt{McLaughlin_1944,McLaughlin_1947,Payne-Gaposchkin_1957,Aydi_etal_2020b}). Typically, the strongest lines in nova spectra are those of the hydrogen Balmer series, 
followed by either low-ionization Fe II lines particularly of the (42, 48, and 49) multiplets, 
or high-excitation He I, He II, N II, and N III lines. These distinct species that dominate nova spectra near the peak brightness 
led to classifying novae under two main spectral classes: ``Fe II'' and ``He/N'' novae \citep{Williams_1992,Williams_2012}. In addition to showing different species of emission/absorption lines, 
the classes are characterized by different ejecta velocities, with Fe II novae showing slower ejecta velocities (FWHM $<$ 2500\,km\,s$^{-1}$) 
compared to He/N novae (FWHM $>$ 2500\,km\,s$^{-1}$) and spectral lines profiles (Fe II novae tend to show more P Cygni line profiles while He/N novae show broad, flat-topped emission lines; \citealt{Williams_1992}).

Mainly, the origin of these spectral features have been attributed to different bodies of gases 
(e.g., \citealt{Williams_2008,Williams_2012}) or changes in the condition of the nova ejecta (e.g., \citealt{Shore_2012,Shore_2013,Shore_2014,Mason_etal_2018}). \citet{Williams_2012} suggested that the Fe II spectra possibly have origin in stripped material from the companion star residing around the binary system, while the He/N spectra originate in ejected material from the surface of the white dwarf. However, \citet{Shore_2014} suggested that the Fe II and He/N spectra are defined by the opacity and ionization state of the ejecta and therefore Fe II novae are observed at a stage where the ejecta are optically thick, while He/N novae are observed after peak at a stage where the ejecta are optically thin (or even completely ionized). 

While novae were assigned specific spectral classes as Fe II or He/N based on their spectra observed after optical peak, \citet{Williams_2012} suggested that if dedicated spectroscopic follow-up were available, 
some if not all novae might show features of both classes simultaneously or evolve from one class to another, and dubbed them as ``hybrid'' novae. 
\citet{Shore_2012,Shore_2014} also suggested that the same nova could show an evolution from one class to another depending on the conditions within the novae ejecta (i.e., changes in opacity, ionization, density). This hybrid evolution has been observed and discussed in the literature for a few novae, e.g., the 2011 eruption of T~Pyx \citep{Shore_etal_2011,Ederoclite_2014,Surina_etal_2014,Arai_etal_2015} and V5558~Sgr \citep{Tanaka_etal_2011}. 

We have entered a new era of all-sky surveys that monitor the sky nightly, facilitating the discovery of novae at the earliest stages of eruption. 
These now-routine early discoveries of novae, combined with rapid spectroscopic follow-up by professional astronomers and citizen scientists
allow us to gather comprehensive spectroscopic data sets for a substantial sample of novae, tracking their eruptions from start to end. Backed by these spectroscopic data sets, we revisit the above mentioned pioneering studies, with the aims of establishing 
a unifying picture of nova spectroscopic evolution and drawing new insights into the origin of the Fe II and He/N spectral classes. 
In Section~\ref{Obs} we discuss our nova sample and the observations and data reduction. In Section~\ref{Res} we present the results, including the spectroscopic evolution of the novae in the sample. These results are further discussed in Section~\ref{Disc}, while our conclusions are presented in Section~\ref{sec_conc}.

\section{Observations and data reduction}
\label{Obs}

Our main sample consists of six well-observed novae, namely T~Pyx (2011), V339~Del (2013), V659~Sct (2019), V1405~Cas (2021), V606~Vul (2021), and Gaia22alz (2022), which all show an analogous spectroscopic evolution throughout their eruptions despite the different timescales of their eruptions. These novae were selected based on the availability of dedicated spectroscopic monitoring throughout the evolution of the nova eruption, from the early rise to peak all the way to the nebular phase. While we are aware that other novae might have comparable dedicated observational follow-up in the literature, we limit our main sample to six novae so we can showcase their evolution clearly in a reasonably compact manuscript. The main sample consists of novae from different speed classes (very fast, fast, slow, and very slow as classified based on the time to decline by two magnitudes from maximum $t_2$; \citealt{Payne-Gaposchkin_1957}). Selecting novae of different speed classes emphasizes that the proposed spectral evolution is common across fast and slow novae. We also show the \textit{early} spectral evolution of four additional, supporting cases: two slow novae (V612~Sct  2017 and FM~Cir 2018); and two very fast novae (V407~Lup 2016 and U~Sco 2022), with the aim of demonstrating the prevalence of the universal spectral evolution we propose in this work, regardless of the speed class of the nova. It is worth noting that all the novae in our sample are known to have dwarf companion stars, with the exception of U~Sco, where the companion star might be more evolved, i.e., a sub-giant donor \citep{Schaefer_1990}. In Table~\ref{table:sample} we list the details of all the novae in our sample.

\begin{table}[!t]
\centering
\caption{The nova sample.}
\begin{tabular}{lcccc}
\hline
Name & $t_0^{(a)}$ & $t_{\mathrm{max}} - t_0$ & $V_{\mathrm{max}}$ & $t_2^{(b)}$\\
 & (UT date) & (days) & (mag) &(days) \\
\hline
\hline
Main Sample  & & & &\\
\hline
T Pyx & 2011-04-14.29 &  28 & 6.2 & 50 $\pm$ 4\\
V339~Del & 2013-08-14.37 & 2 & 4.5 & 11  $\pm$ 2\\
V659 Sct &2019-09-29.06 & 2.0 & 8.3 & 7 $\pm$ 1 \\
V1405~Cas & 2021-03-18.42 & 53 & 5.1 & 165 $\pm$ 5\\
V606~Vul & 2021-07-16.49 & 16.5 & 10.0 & 87 $\pm$ 4\\
Gaia22alz & 2022-01-25.02 & 178 & 10.8 & 207 $\pm$ 5\\
\hline
Slow novae & & & &\\
\hline
V612~Sct & 2017-06-19.50 & 40 & 8.4 & $>$ 172\\
FM~Cir & 2018-01-17.87 & 25 & 6.5 & 120  $\pm$ 5\\
\hline
Fast novae & & & &\\
\hline
V407~Lup & 2016-09-24.00 & 1.4 & 5.6 & $\leq$ 2.9\\
U~Sco & 2022-06-06.72 & 0.5 & 7.7 & 2 $\pm$ 0.5\\
\hline
\end{tabular}
{\raggedright $^{a}$ $t_0$ is the discovery date.\\
$^{b}$ $t_2$ is derived as the time between peak brightness and the last time the nova drops by 2\,mag. from peak.}
\label{table:sample}
\end{table}

The spectra presented in this paper derive from a variety of facilities and sources, including both professional observatories and contributions from citizen scientists. Here we provide a brief summary of the facilities and instruments utilized for spectroscopic observations, while the detailed observation logs for each nova may be found in Tables~\ref{table:spec_log_T_Pyx},~\ref{table:spec_log_V339_Del},~\ref{table:spec_log_V659_Sct},~\ref{table:spec_log_V1405_Cas},~\ref{table:spec_log_V606_Vul},~\ref{table:spec_log_Gaia22alz},~\ref{table:spec_log_V612_Sct},~\ref{table:spec_log_FM_Cir},~\ref{table:spec_log_V407_Lup}, and ~\ref{table:spec_log_U_Sco} .

\begin{itemize}

\item A fraction of the spectra used in this work were obtained using the Goodman spectrograph \citep{Clemens_etal_2004} on the 4.1\,m Southern Astrophysical Research (SOAR) telescope located on Cerro Pach\'on, Chile. The spectra were reduced and optimally extracted using the \textsc{apall} package in IRAF \citep{Tody_1986}.

\item Another fraction of the spectra used in this paper were obtained using the High Resolution Spectrograph (HRS; \citealt{Barnes_etal_2008,Bramall_etal_2010,Bramall_etal_2012,Crause_etal_2014}) and the Robert Stobie Spectrograph (RSS; \citealt{Burgh_etal_2003}; \citealt{Kobulnicky_etal_2003}) mounted on the Southern African Large Telescope (SALT; \citealt{Buckley_etal_2006,Odonoghue_etal_2006}) in Sutherland, South Africa. The primary reduction of the HRS spectroscopy was conducted using the SALT science pipeline \citep{Crawford_etal_2010}, which includes over-scan correction, bias subtraction, and gain correction. The rest of the reduction was done using the MIDAS HRS pipeline described in details in \citet{kniazev_etal_2016,2019AstBu..74..208K}.

\item We also obtained spectroscopic data using the ESO-VLT 8-m telescope with the Fibre-fed Optical Echelle Spectrograph (FEROS; \citealt{Kaufer_etal_1999}). The spectra were reduced and optimally extracted 
using the \textsc{apall} package in IRAF. 

\item Spectral observations were also gathered with the Asiago 1.22m + B\&C telescope. The spectra were fully reduced and fluxed against nightly standards in IRAF following the procedures detailed in the \citet{2000iasd.book.....Z}
cookbook.

\item We also made use of publicly available data from the Astronomical Ring for Access to
Spectroscopy (ARAS; \citealt{Teyssier_2019}). These consists of a combination of low-resolution ($R\approx 1000)$ and medium-resolution (up to $R\approx 11000)$ spectra obtained by citizen scientists. 

\end{itemize}

We also make use of publicly available photometry from 
the American Association of Variable Stars (AAVSO; \citealt{AAVSODATA}) International Database and the All-Sky automated survey for Supernovae (ASAS-SN; \citealt{Shappee_etal_2014}). These data consists of CCD and CMOS photometry, mostly in the $V$-band, $g$-band,
and unfiltered band with $V$ zero-point ($CV$), 
as well as visual estimates; these measurements are used to build the optical light curves for the novae in our sample (Figures~\ref{Fig:T_Pyx_main_spec}, \ref{Fig:V339_Del_main_spec}, \ref{Fig:V659_Sct_main_spec}, \ref{Fig:V1405_Sct_main_spec_1}, \ref{Fig:V606_Vul_main_spec}, \ref{Fig:Gaia22alz_main_spec}, \ref{Fig:V612_Sct_main_spec}, \ref{Fig:FM_Cir_main_spec}, \ref{Fig:V407_Lup_main_spec}, and~\ref{Fig:U_Sco_main_spec}).

\section{Results}
\label{Res}

In this section we present the spectral evolution of all the novae in our sample, dividing them into two groups: the main sample consisting of six novae and the supporting sample (consisting of two slow and two very fast novae). 
Here we present the spectroscopic evolution of our nova sample illustrated by selected spectroscopic epochs, while the complete spectroscopic evolution are presented in supplementary online material. 

\subsection{The main sample}

Our main sample consists of six well-observed novae, which show comparable spectroscopic evolution, evolving through at least three distinct phases before they reach the nebular phase. The phases described below are as follow:
$$\mathrm{Early\,\,He/N\,\,(1)} \longrightarrow \mathrm{Fe~II\,\,(2)} \longrightarrow \mathrm{Late\,\,He/N\,\,(3)}\,.$$ 

Below we describe the spectroscopic evolution of each nova in our main sample. 
\\
\\
\noindent \textbf{Nova T~Pyx (2011):} the 2011 eruption of the recurrent Galactic nova T~Pyxidis was discovered by M. Linnolt
on 2011 Apr 14.29 UT (HJD 2455665.79; which will be considered as $t_0$ for this nova; see \citealt{Schaefer_etal_2013} for more details). \citet{Surina_etal_2014} and  \citet{Arai_etal_2015} presented detailed spectrophotometric studies of the 2011 eruption of T~Pyx, focusing on the changes in the spectra and the development of the absorption/emission lines. 

\begin{figure*}
\begin{center}
  \includegraphics[width=0.62\textwidth]{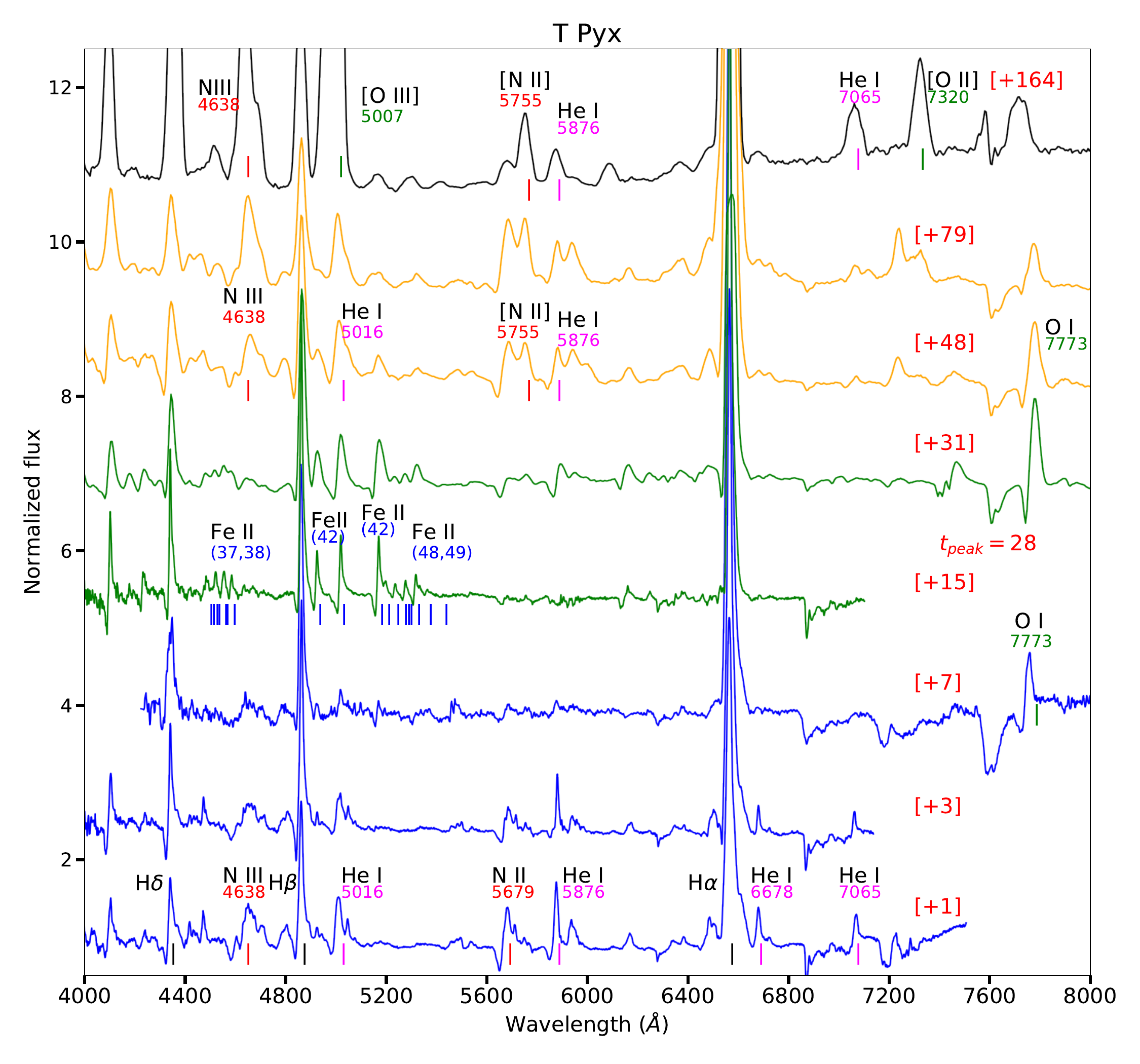}\includegraphics[width=0.388\textwidth]{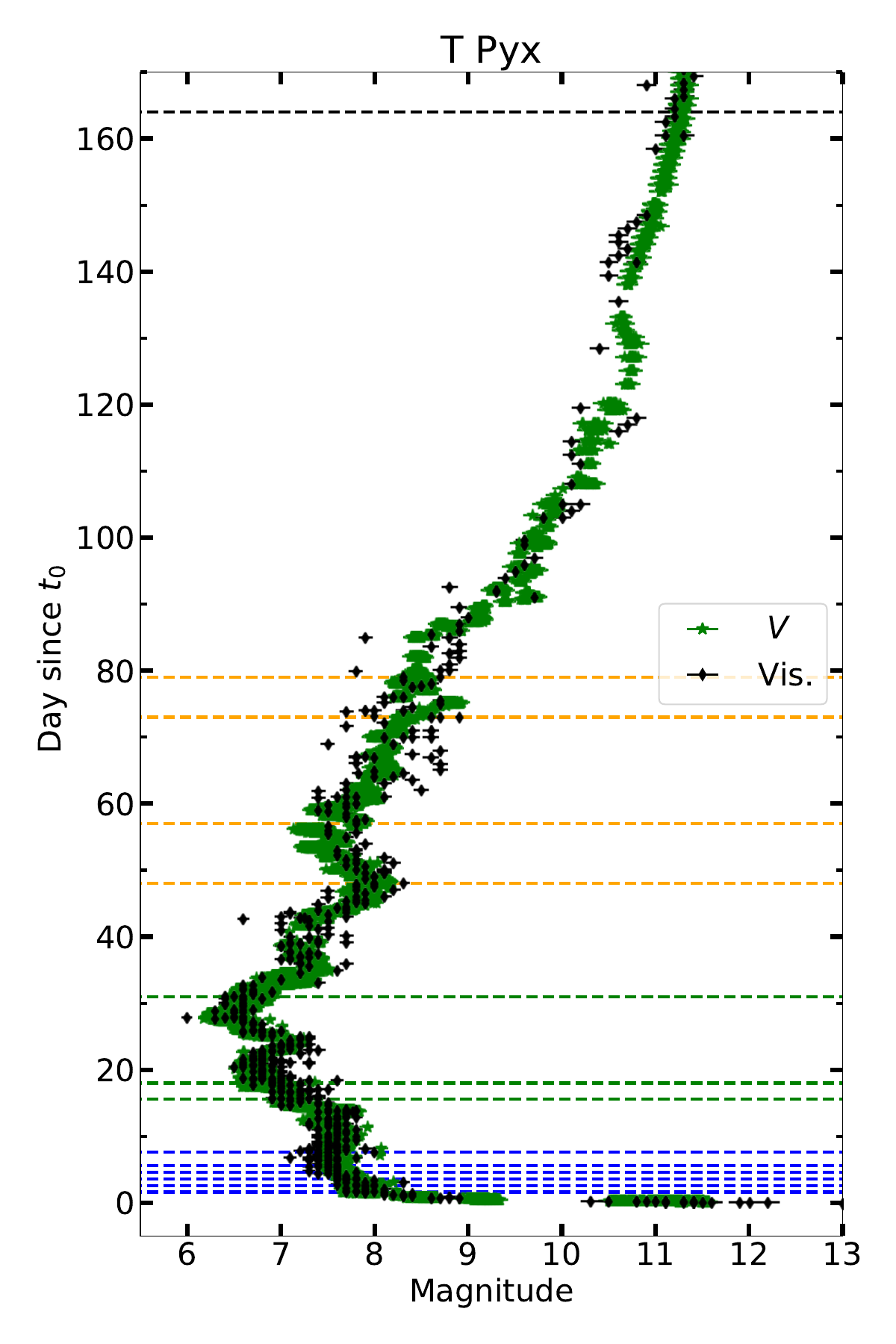}
\caption{\textit{Left:} the overall spectroscopic evolution of nova T Pyx representing the different spectral stages:  phase 1 (early He/N), highlighted in blue); phase 2 (Fe II), highlighted in green; phase 3 (late He/N), highlighted in orange; and the nebular phase, highlighted in black. Numbers between brackets are days after $t_0$. Tick marks are presented under the lines for easier identification and they are color coded based on the line species. \textit{Right:} the optical light curve of nova T Pyx. The blue, green, orange, and black dashed lines represent the spectroscopic epochs for when the nova was in phase 1 (early He/N), phase 2 (Fe II), phase 3 (late He/N), and the nebular phase, respectively.}
\label{Fig:T_Pyx_main_spec}
\end{center}
\end{figure*}

In Figure~\ref{Fig:T_Pyx_main_spec} we present the optical light curve of the nova, which belongs to a moderately fast speed class ($t_2 = 50$ days). It took 28 days for the nova to reach optical peak from the time of discovery. Spectroscopic monitoring starting from day 1.6 all the way to day 164 shows that the nova goes through three phases before it enters the nebular phase. These phases are plotted in Figure~\ref{Fig:T_Pyx_main_spec}, where the spectra are highlighted in different colors for clarity. The spectroscopic evolution consists of: phase 1, highlighted in blue, when the spectra are dominated by P Cygni profiles of Balmer, He I, and N II, and N III, lasting till day 8 after $t_0$. This phase, covers the early rise to visible peak; phase 2, highlighted in green, when the spectra are dominated by P Cygni profiles or emission lines of Balmer and Fe II and covering the late rise and early decline from visible peak (between days 15 and 40); phase 3, highlighted in orange, when the spectra are again dominated by high-excitation lines of He I, He II, N II, N III, along with Balmer lines. For T Pyx, this phase extends from day 40 until the nova enters the nebular phase around day 150. As the nova enters the nebular phase (highlighted in black in Figure ~\ref{Fig:T_Pyx_main_spec}), nebular forbidden lines of singly and doubly ionized oxygen emerge and eventually high-ionization emission lines of forbidden Fe. 
A detailed spectroscopic evolution of T~Pyx, including more spectral epochs is available in the online supplementary material.\\ 

\begin{figure*}
\begin{center}

\includegraphics[width=0.62\textwidth]{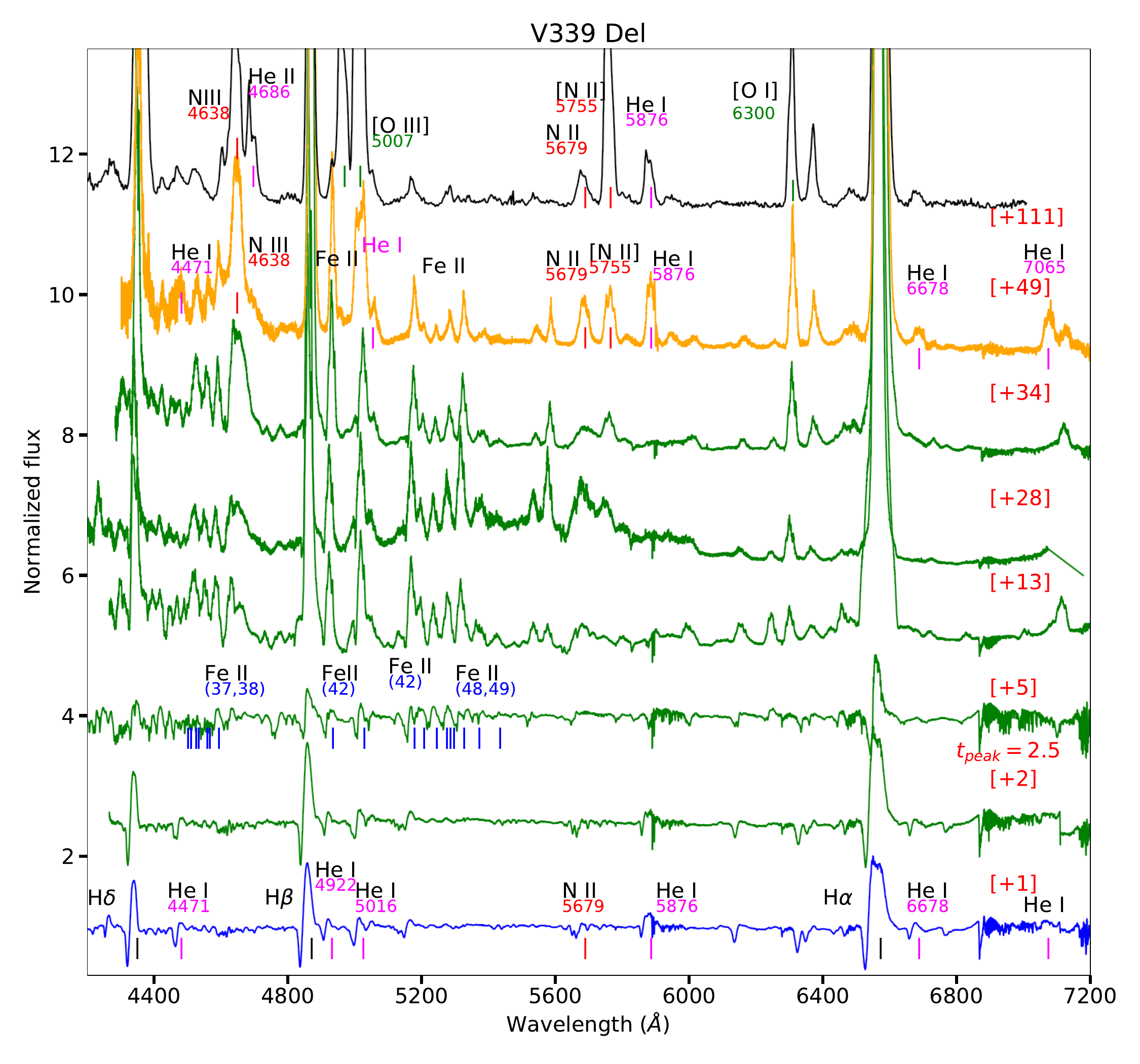}\includegraphics[width=0.388\textwidth]{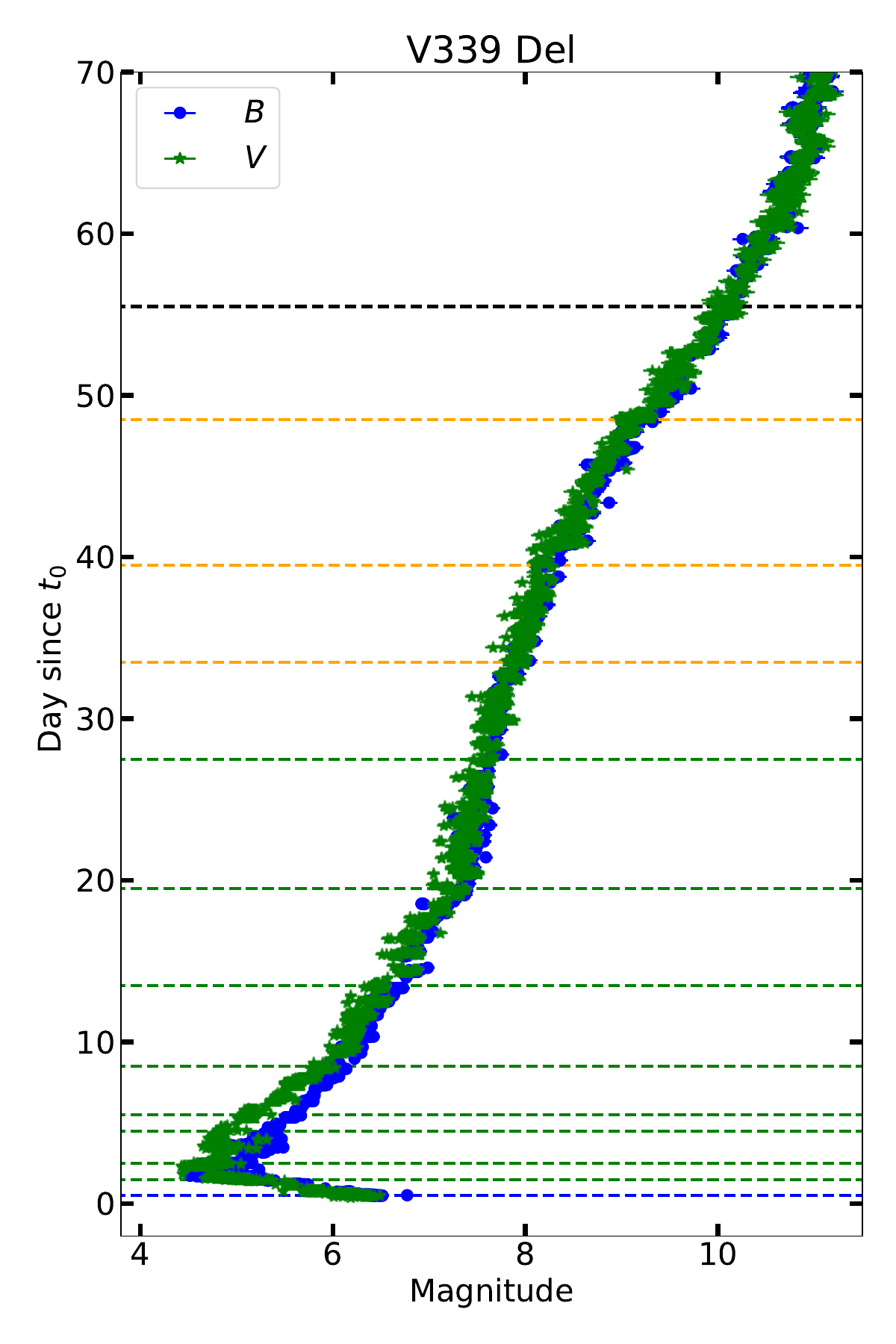}
\caption{Same as Figure~\ref{Fig:T_Pyx_main_spec} but for nova V339~Del.}
\label{Fig:V339_Del_main_spec}
\end{center}
\end{figure*}

\noindent \textbf{Nova V339~Del:}
the fast nova V339~Del was discovered by Koichi Itagaki on 2013 Aug 14.584 UT at a visual magnitude of around 6.8 \citep{2013CBET.3628....1N}. Pre-discovery observations of the nova show that the eruption had started as early as 2013 Aug 14.37 (HJD 2456518.90 = $t_0$; \citealt{ATel_5316}). The overall spectroscopic evolution of the nova and its optical light curve are presented in Figure~\ref{Fig:V339_Del_main_spec}. The nova evolution was fast, with $t_2$ = 11 days. It also took just 2 days for the nova to reach optical peak from the time of discovery. The first spectrum obtained for the nova, less than a day after discovery, shows P Cygni lines of H Balmer and He I, with relatively weak Fe II P Cygni lines. In the following epochs, taken over the following days, the He I lines weakened, while the Fe II lines became more prominent. Due to the presence of weak Fe II lines in the first epoch, we suggest that this epoch was obtained during a relatively rapid (a few hours duration) phase 1 (early He/N) or during a transition between phase 1 (early He/N) and phase 2 (Fe II). The Fe II lines dominated the spectrum until day 28 after $t_0$. After this, emission lines of He I, N III, and [N II] emerged and strengthened gradually until day around 50. Thereafter, strong emission lines of [O III], [O II], [N II], and He II emerged, indicating that the nova had entered the nebular phase. A detailed spectroscopic evolution of nova V339~Del is available in the supplementary online material. 

\begin{figure*}
\begin{center}
  \includegraphics[width=0.62\textwidth]{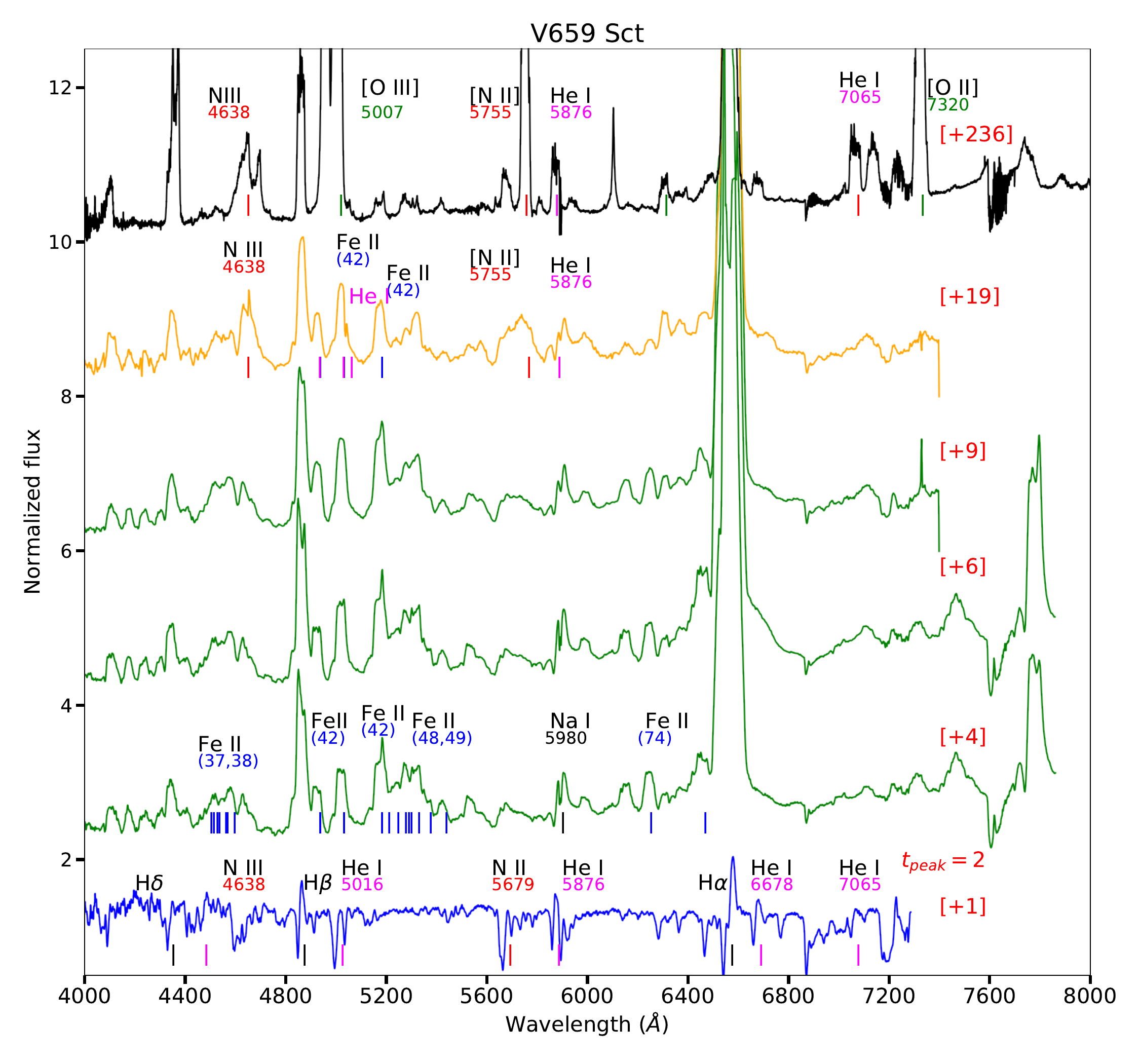}\includegraphics[width=0.388\textwidth]{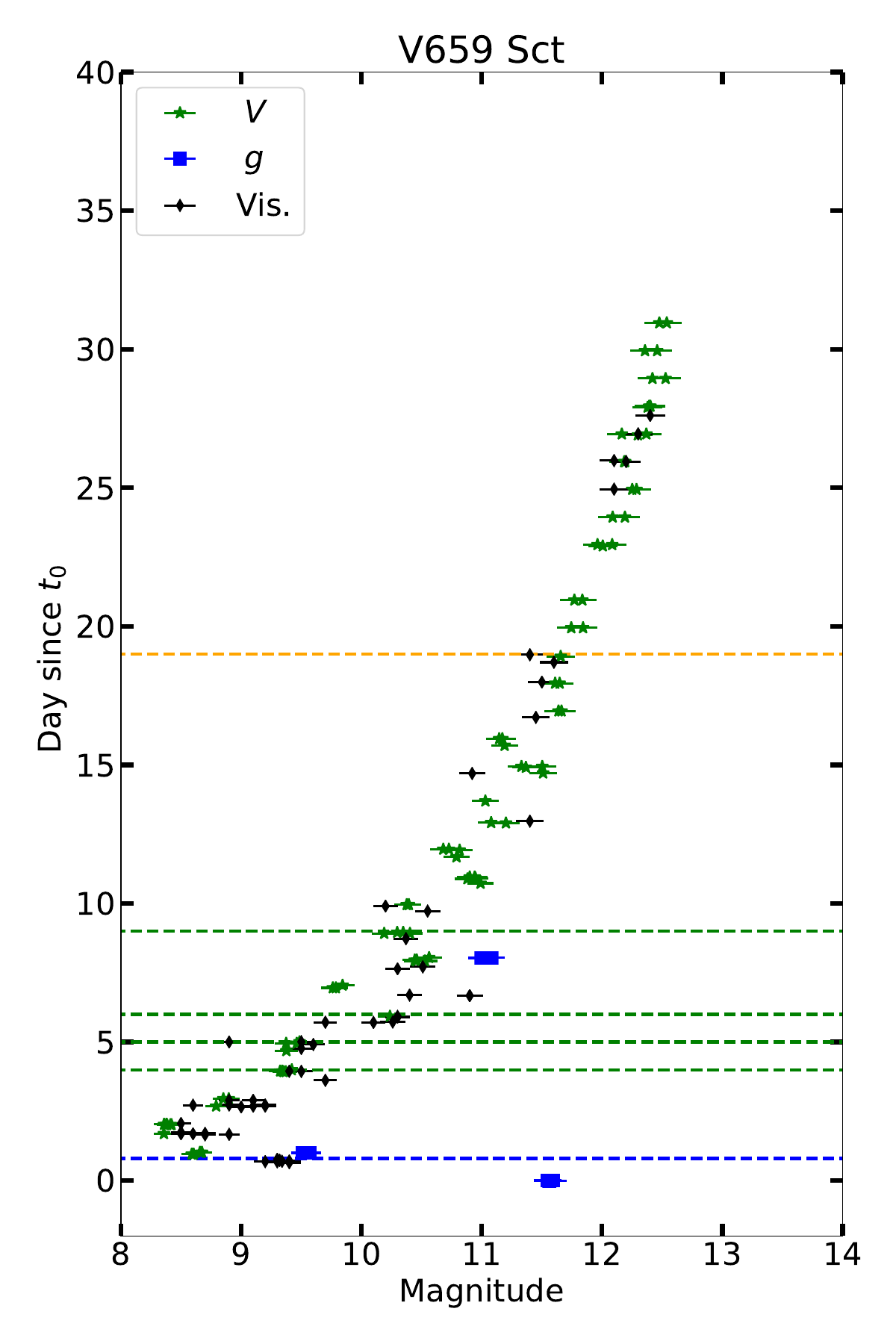}
\caption{Same as Figure~\ref{Fig:T_Pyx_main_spec} but for nova V659~Sct.}
\label{Fig:V659_Sct_main_spec}
\end{center}
\end{figure*}

\noindent \textbf{Nova V659~Sct (ASASSN-19aad):} the very fast Galactic nova V659~Sct was discovered by ASAS-SN on 2019 Oct 29.05 UT (HJD 2458785.55 = $t_0$), as ASASSN-19aad \citep{2019TNSTR2216....1S}. The overall spectroscopic evolution of the nova and its optical light curve are presented in Figure~\ref{Fig:V659_Sct_main_spec}. The nova evolution was very fast, with $t_2$ = 7 days, while it reached optical peak in 2 days only since $t_0$. The first spectrum obtained for the nova, less than a day after discovery, showed P Cygni profiles of H Balmer, He I, and N III. A few days later, the He and N lines weakened, while emission lines of Fe II and Na I dominated the spectrum. By day 19, the spectrum is again dominated by a combination of He I, N III, [N II] emission lines, along with Fe II and [O I]. The nova went into solar conjunction afterwards. spectroscopic observations taken over 200 days from $t_0$ showed strong forbidden lines of oxygen and nitrogen ([O III], [O II], and [N II]), implying that the nova has entered the nebular phase  (Figure~\ref{Fig:V659_Sct_main_spec}).
\\

\begin{figure*}
\begin{center}
  \includegraphics[width=\columnwidth]{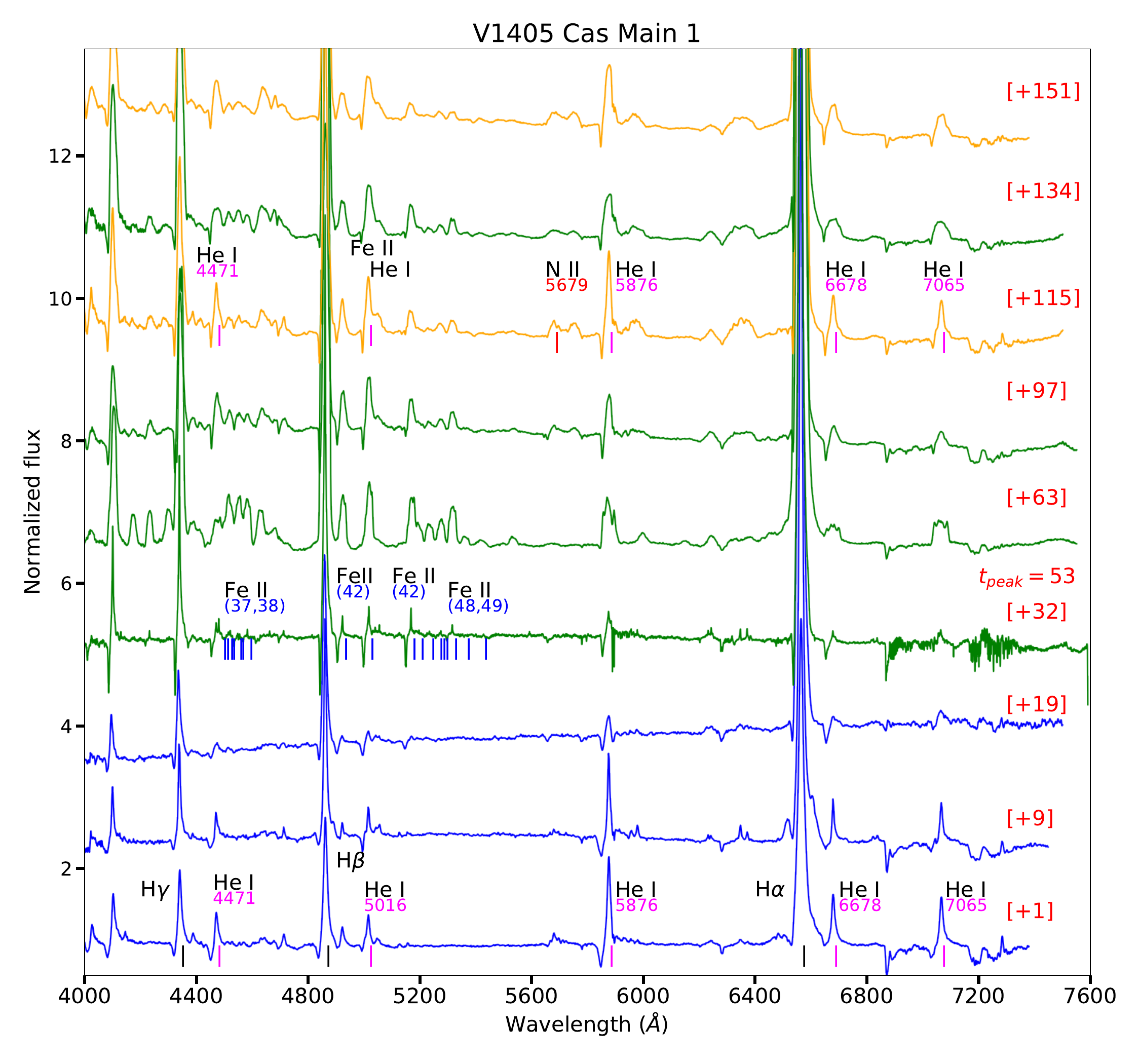}
\includegraphics[width=\columnwidth]{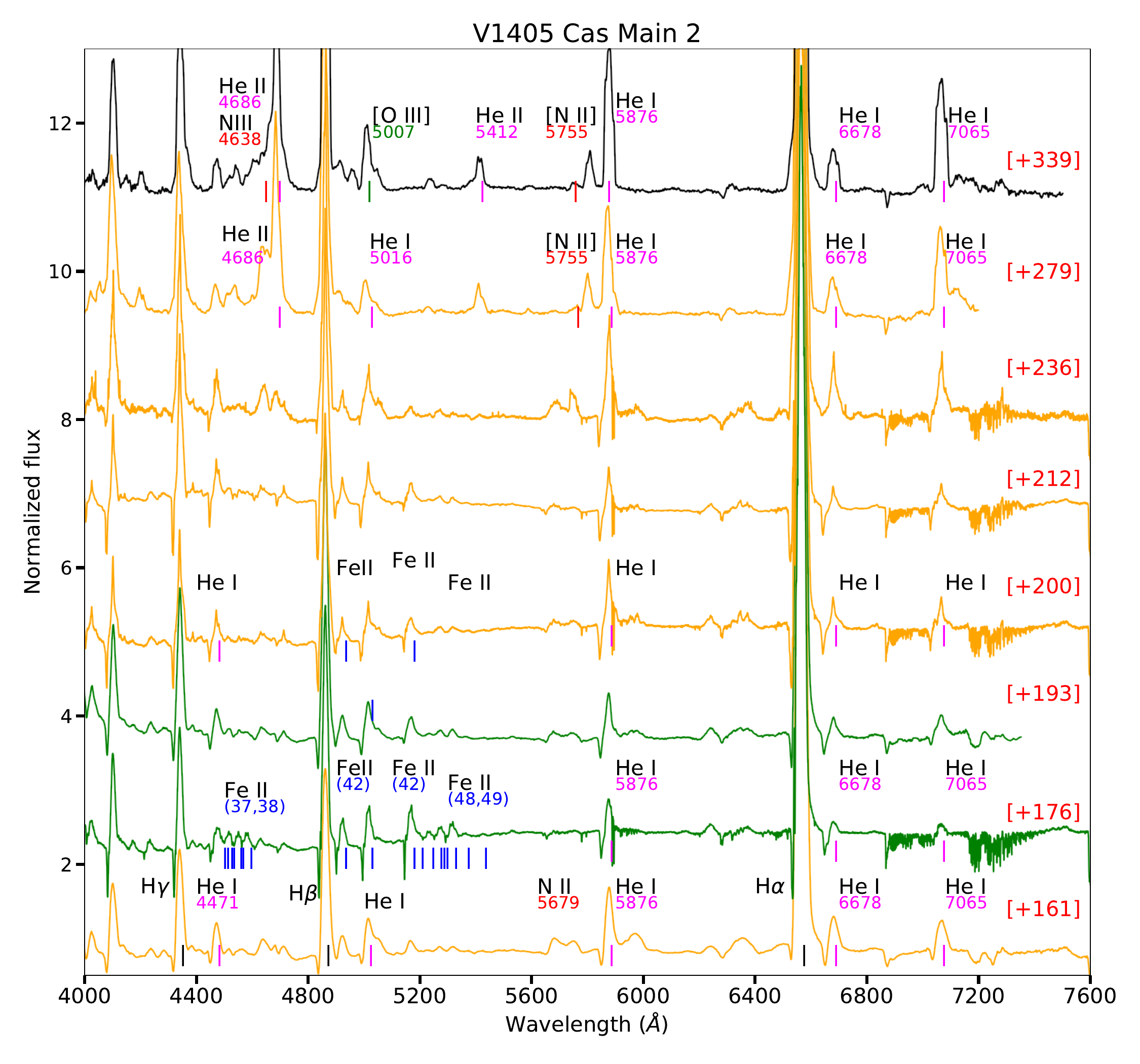}
\includegraphics[width=0.75\textwidth, height = 9cm]{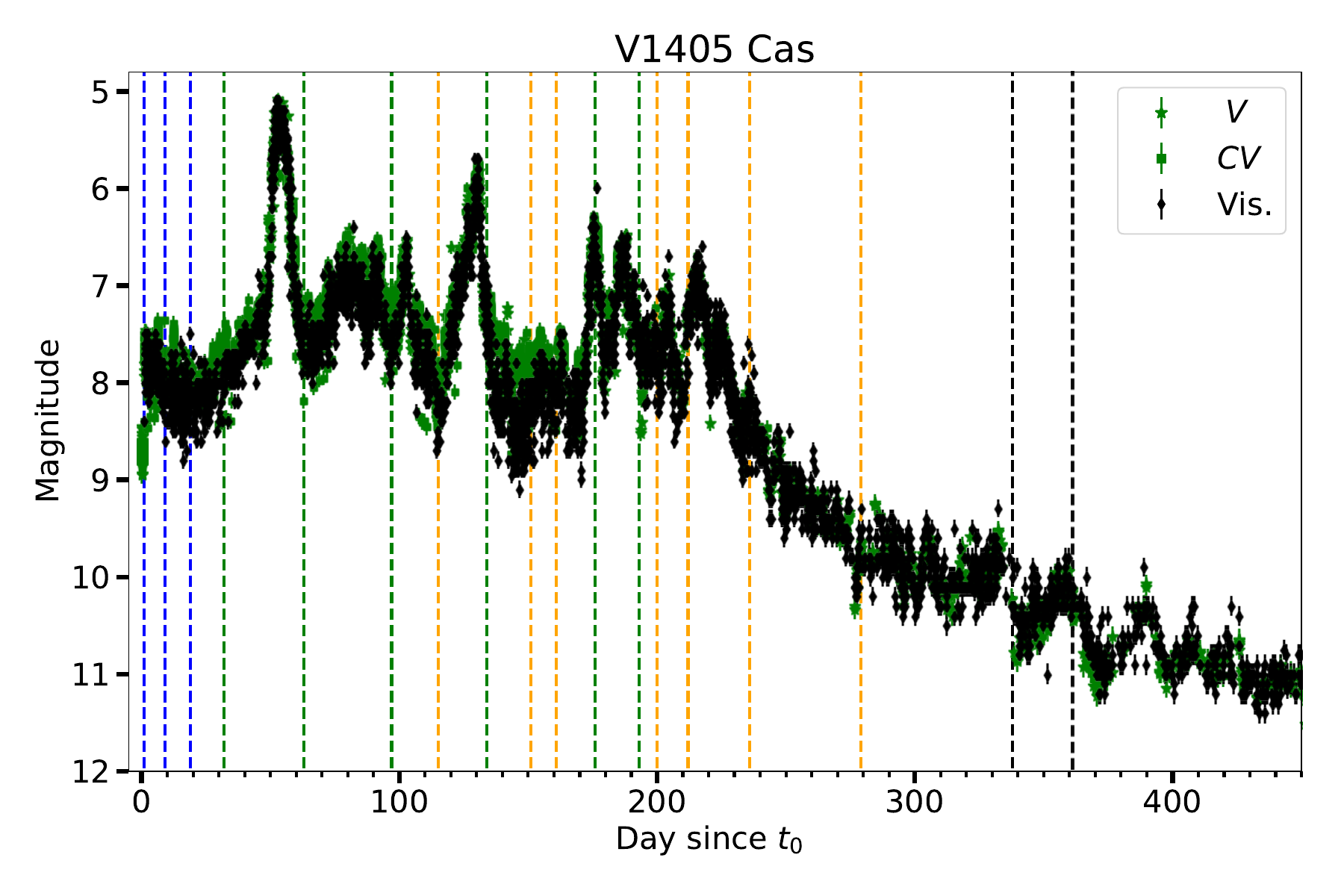}
\caption{Same as Figure~\ref{Fig:T_Pyx_main_spec} but for nova V1405 Cas.}
\label{Fig:V1405_Sct_main_spec_1}
\end{center}
\end{figure*}

\noindent \textbf{Nova V1405~Cas:} the very slow Galactic classical nova V1405~Cas was discovered by Yuji Nakamura on 2021 Mar 18.42 UT (HJD 2459291.92 = $t_0$). The overall spectroscopic evolution of the nova and its optical light curve are presented in Figure~\ref{Fig:V1405_Sct_main_spec_1}. The nova took 53 days to reach the first visible peak. It also took more than 165 days ($t_2$) to decline by 2 mag from peak, as it stayed near visible peak for months, showing multiple flares (maxima). There are two major maxima/flares peaking on day $\approx$ 53 and day $\approx$ 130, at $V= 5.1$ and $V= 5.7$, respectively. The light curve also showed multiple smaller-amplitude flares peaking on days around 79, 92, 103, 175, 187, 205, and 2016. Between days 1 and 19 after $t_0$, the spectra of V1405~Cas were dominated by P Cygni profiles of Balmer and He I lines. From day 32 up to day around 100, the He I lines weakened significantly and Fe II P Cygni lines of the (42, 48, and 49) multiplets emerged and dominated the spectra (apart from the Balmer lines). Thereafter, the nova showed an oscillatory behavior where at some epochs the Fe II lines were dominant, while in other epochs the He I lines were the dominant ones. This oscillatory behavior is accompanied with the appearance of flares in the optical the light curve (Figure~\ref{Fig:V1405_Sct_main_spec_1}). We further elaborate on this oscillatory behavior in Section~\ref{Disc}. After day 340, [O III] nebular lines emerge, implying that the nova has approached the nebular phase.\\

\begin{figure*}
\begin{center}
  \includegraphics[width=0.62\textwidth]{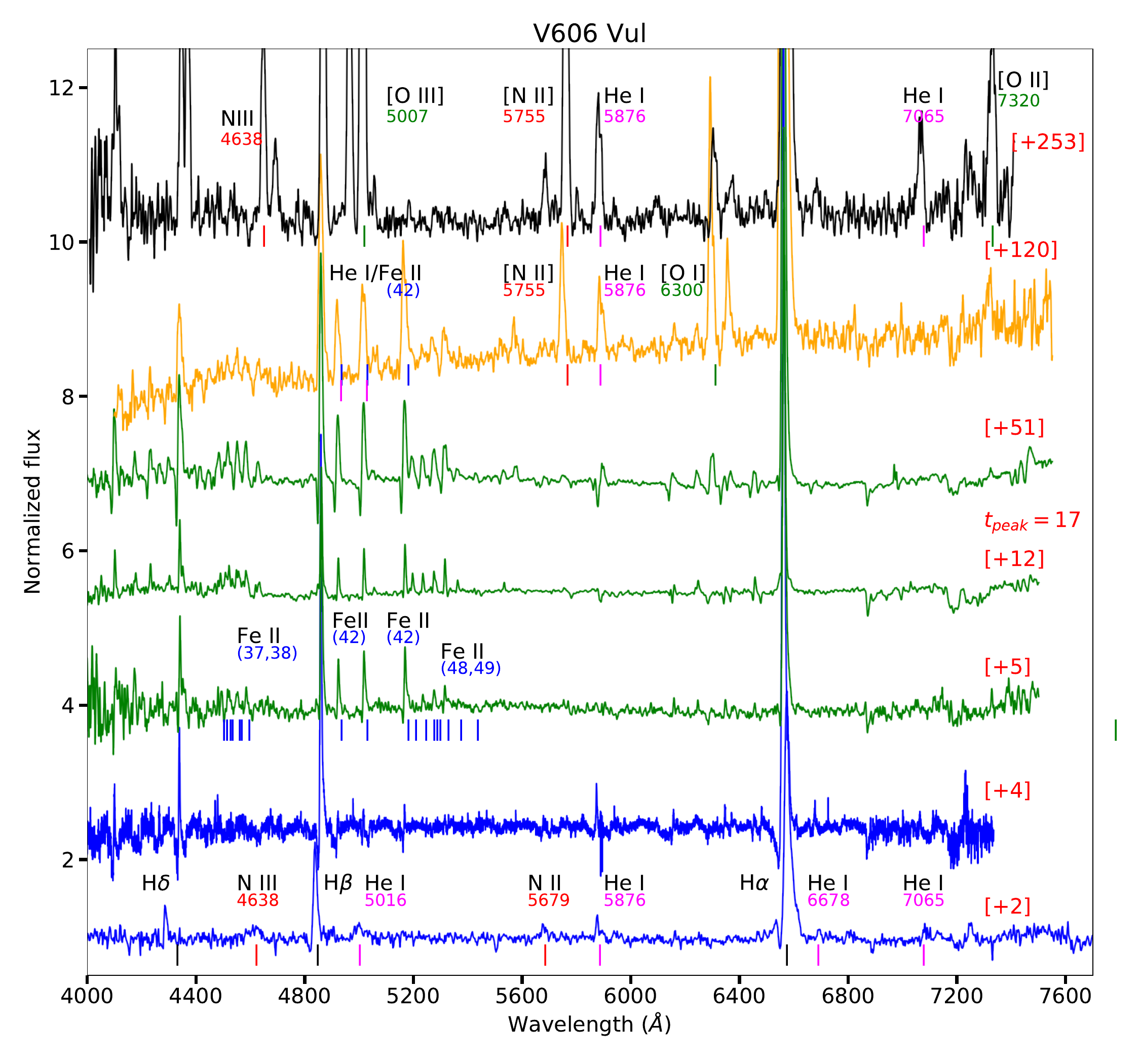}\includegraphics[width=0.388\textwidth]{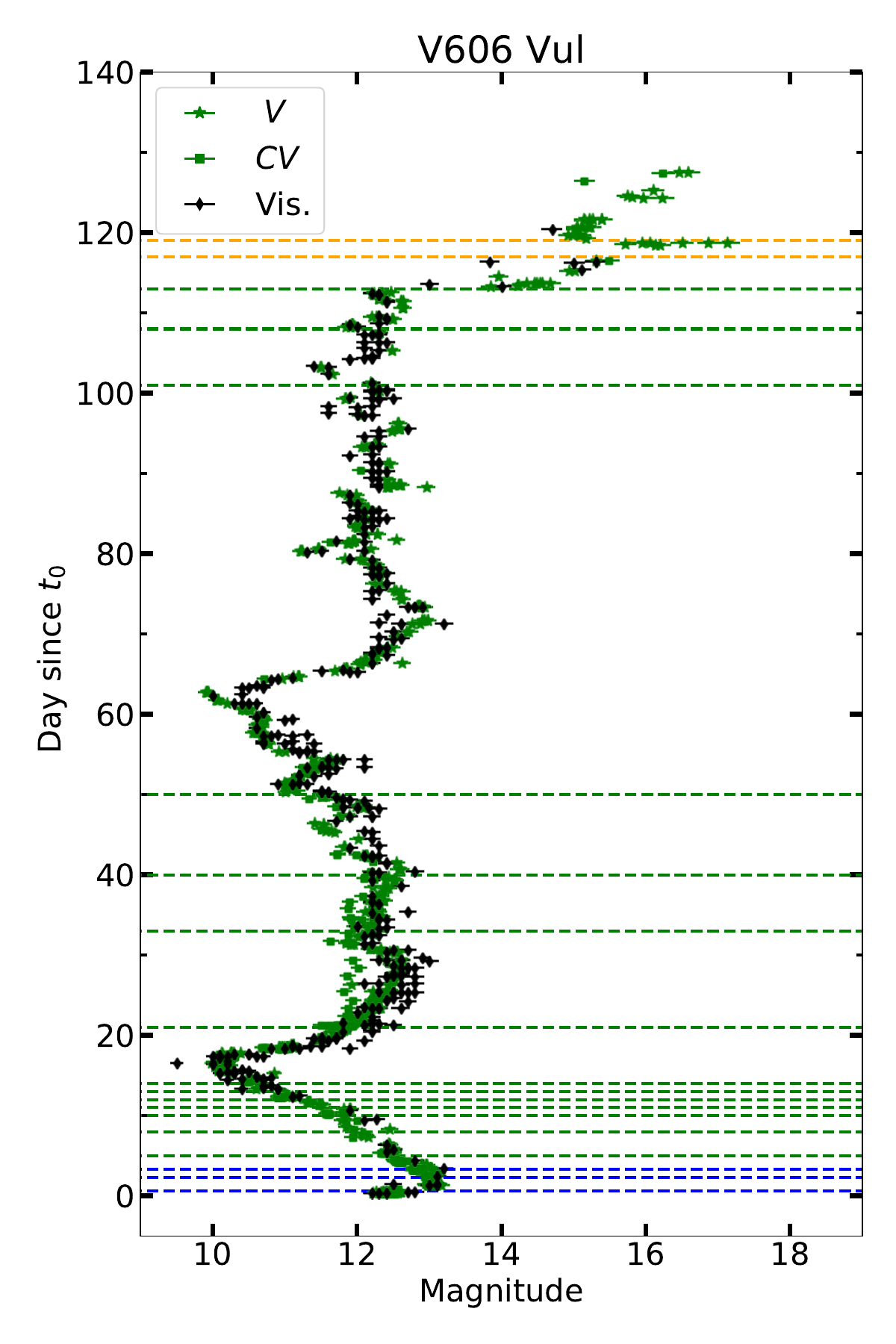}
\caption{Same as Figure~\ref{Fig:T_Pyx_main_spec} but for nova V606~Vul.}
\label{Fig:V606_Vul_main_spec}
\end{center}
\end{figure*}

\noindent \textbf{Nova V606~Vul:} the slow Galactic classical nova V606~Vul was discovered by Koichi Itagaki on 2021 Jul 16.47~UT (HJD 2459411.97). Pre-discovery observations by the Zwicky Transient Facility (ZTF; \citealt{Masci_etal_2019}) show that the eruption started by 2021 Jun 15.38~UT (HJD 2459410.88 = $t_0$; Sokolovsky et al. in prep.). The overall spectroscopic evolution of the nova and its optical light curve are presented in Figure~\ref{Fig:V606_Vul_main_spec}. The nova evolution was slow, with $t_2$ = 87 days, while it climbed to optical peak in 16 days. The optical light curve is characterized by two main flares (maxima), peaking on days around 16 and 63 (at $V \approx$ 10), with several other smaller-amplitude flares across the first 4 months of the eruption. Similar to the spectral evolution of the other novae in our main sample, V606~Vul showed initially spectra dominated by P Cygni profiles of Balmer, He I, N II, and N III during the early rise to peak between days 1 and 3. Afterwards, the He and N lines faded while Fe II lines of the 42, 48, and 49 multiplets emerged (see Figure~\ref{Fig:V606_Vul_main_spec}). This phase lasted between days 5 and 113, before the Fe II lines weakened and high-excitation lines of He and N (permitted and forbidden) strengthened in the following epochs (days 117 and 119). This was followed by a period of solar conjunction where the system was not observable. After V606~Vul emerged from solar conjunction, strong emission lines of forbidden, single and double ionized oxygen and nitrogen were in the spectrum, meaning that the nova already entered the nebular phase. A detailed spectroscopic evolution of nova V606~Vul is available in the supplementary online material.\\  

\begin{figure*}
\begin{center}
  \includegraphics[width=0.62\textwidth]{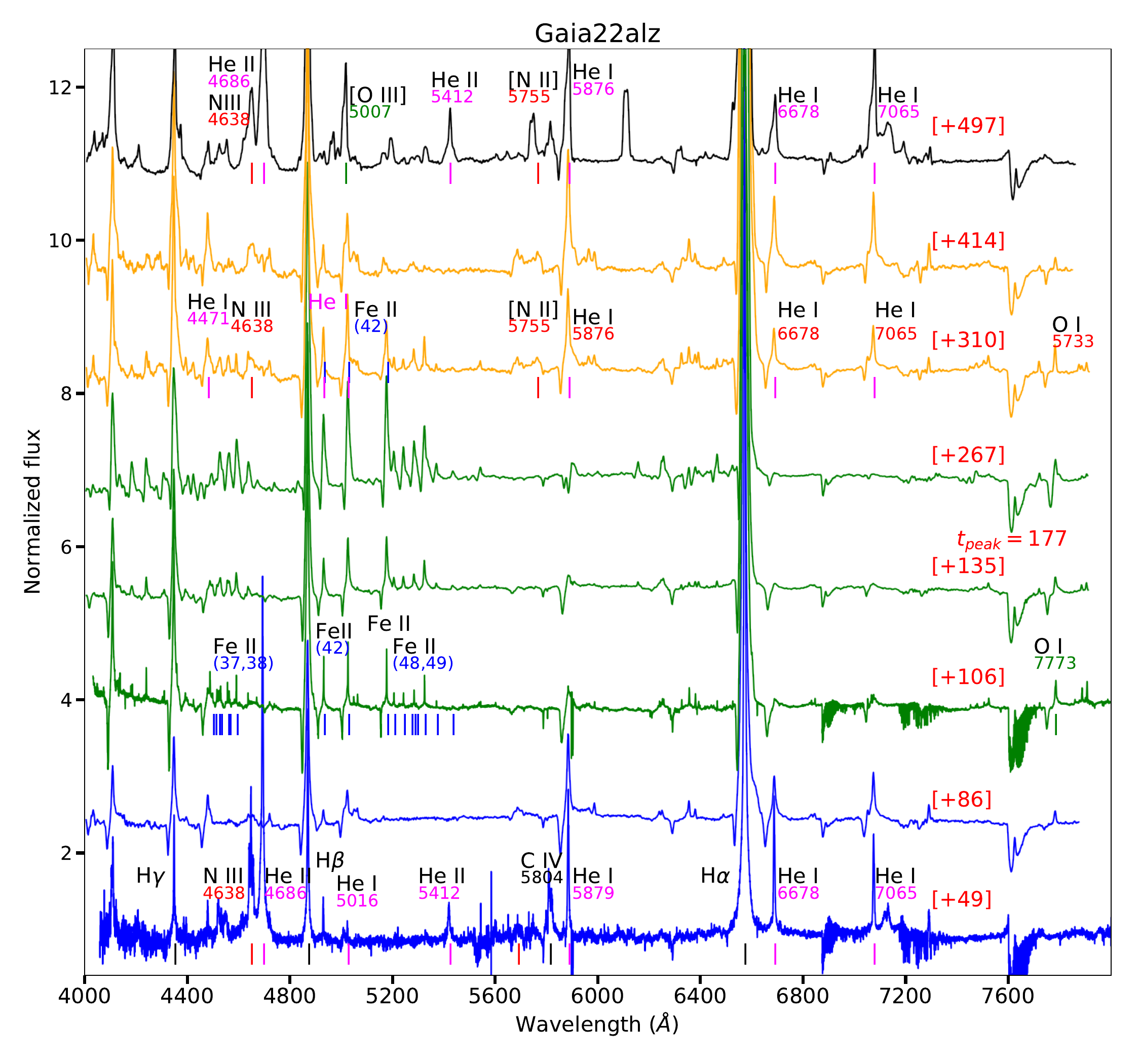}\includegraphics[width=0.388\textwidth]{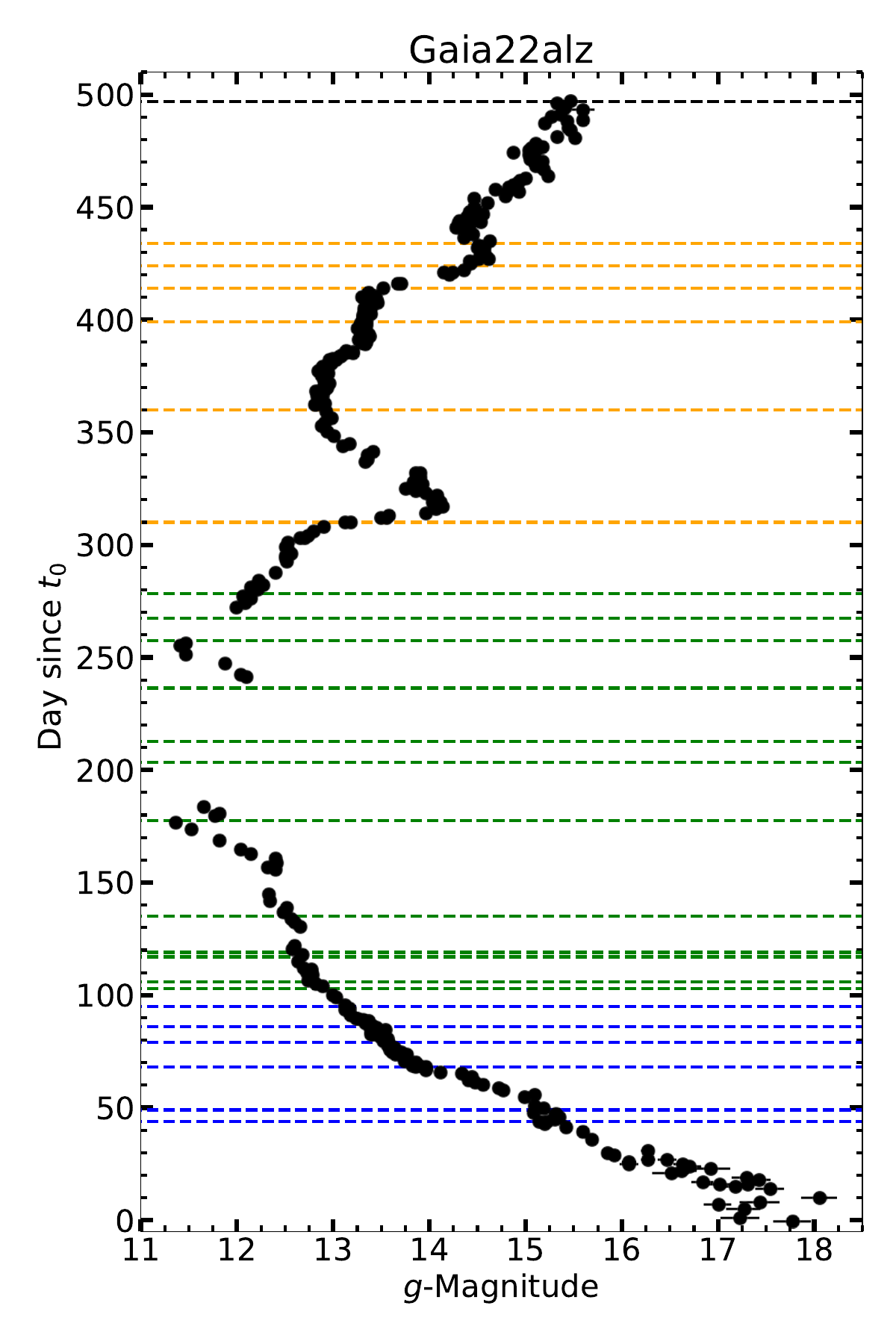}
\caption{Same as Figure~\ref{Fig:T_Pyx_main_spec} but for nova Gaia22alz.}
\label{Fig:Gaia22alz_main_spec}
\end{center}
\end{figure*}

\noindent \textbf{Gaia22alz:} the very slow Galactic nova Gaia22alz was reported on 2022 Feb 04.25 UT by Gaia alerts as an optical Galactic transient \citep{2022TNSTR.313....1H}. Pre-discovery photometry from ASAS-SN showed that the eruption started as early as 2022 Jan 25.02 UT (HJD 2459604.52 = $t_0$; see \citealt{Aydi_etal_2023a} for more details). The overall spectroscopic evolution of nova Gaia22alz and its optical light curve are presented in Figure~\ref{Fig:Gaia22alz_main_spec}. The nova evolution was very slow, with $t_2$ = 207 days. It also took 178 days for the nova to reach optical peak from $t_0$, making it the slowest rising nova in our sample. The first few epochs of this nova were dominated by emission lines and later P Cygni profiles, of Balmer, He I, N II, and N III between days 45 and 97. A detailed description of the early spectroscopic evolution of this nova is presented in \citet{Aydi_etal_2023a}. The He and N lines weakened and the Fe II lines of multiplets (42, 48, and 49) strengthened, as the nova climbed to optical peak between days 103 and 278. From day 310 onwards, the Fe II lines became less prominent, while He I, He II, [N II], N III emerged and dominated the spectra (in addition to Balmer lines). By day 497, we observe relatively strong [O III] line at 5007 \AA, implying that the nova has approached the nebular phase. A detailed spectroscopic evolution of nova Gaia22alz is available in the supplementary online material. 

\subsection{Slow novae (traditionally Fe II)}
In this section we present the early spectroscopic evolution of two novae (a slow and very slow one), which are traditionally classified as Fe II class. These novae show comparable behavior to the novae above, but lack coverage for all four phases. We focus on their early spectroscopic evolution, demonstrating that they show evidence for early He/N spectral features (phase 1), if caught early enough. Below we describe these two slow novae: 
\\

\begin{figure*}
\begin{center}
  \includegraphics[width=0.62\textwidth]{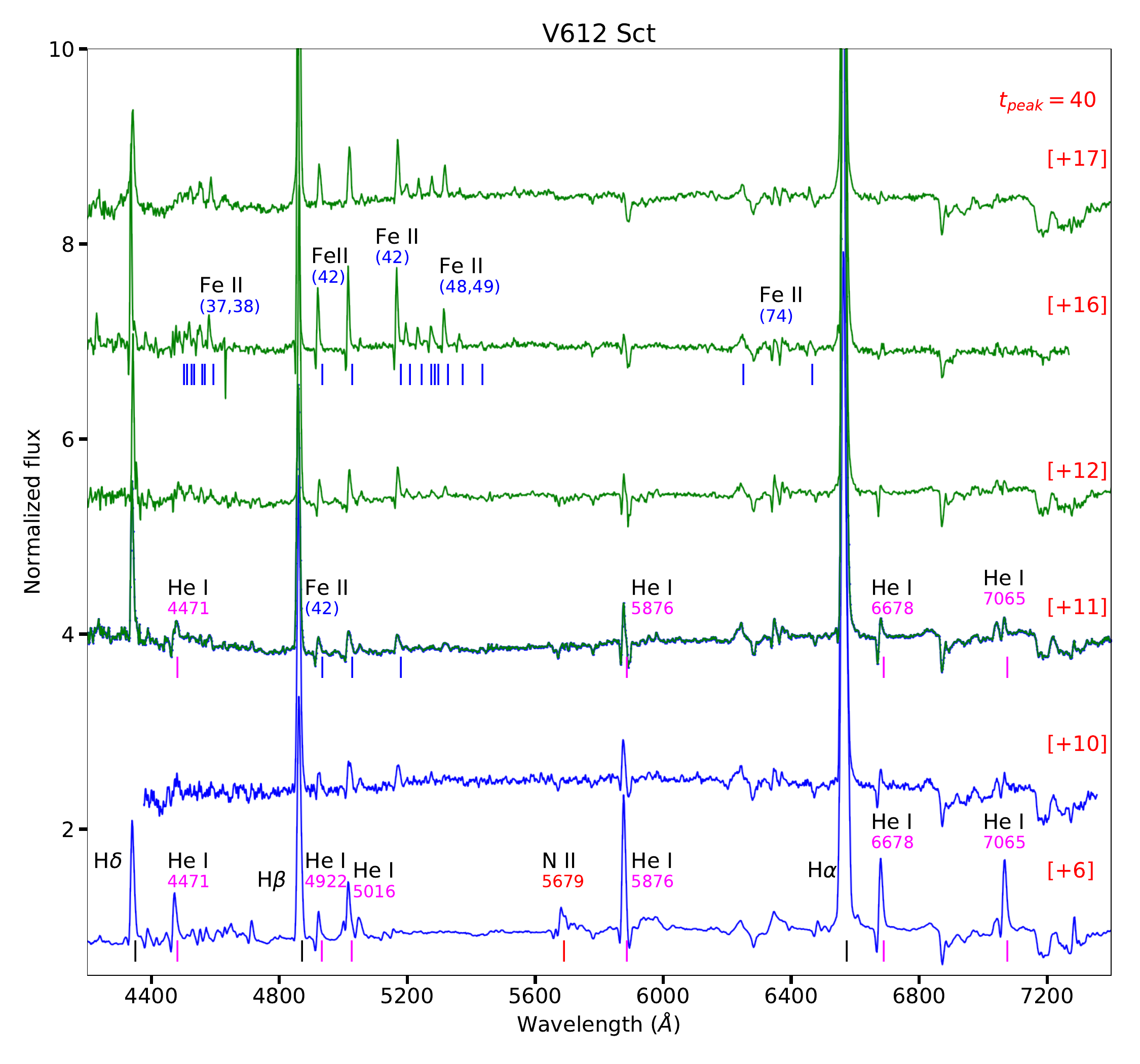}\includegraphics[width=0.388\textwidth]{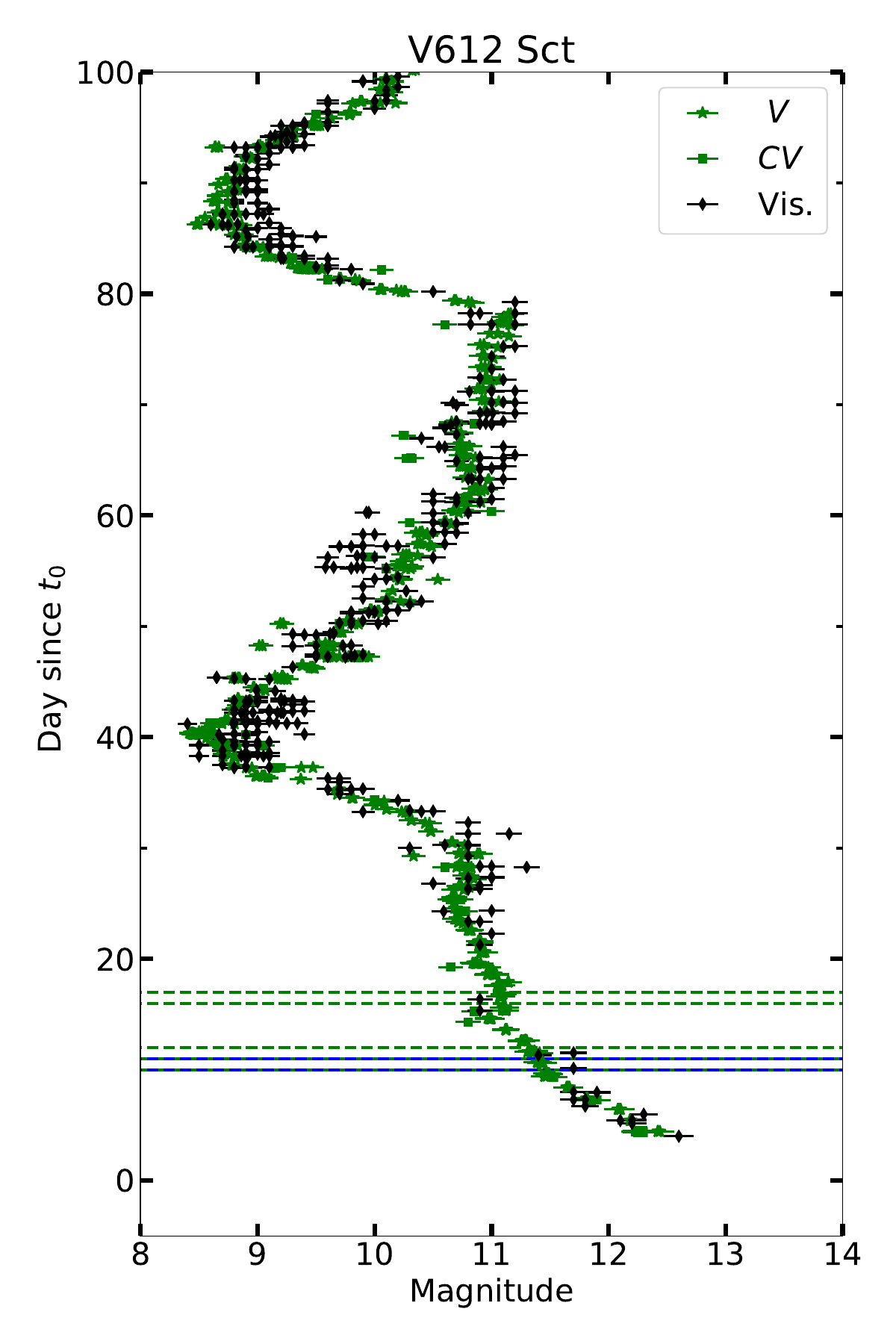}
\caption{Same as Figure~\ref{Fig:T_Pyx_main_spec} but for nova V612~Sct. The spectrum on day 11 is plotted in both green and blue, indicating that this epoch is a transition from phase 1 to phase 2.}
\label{Fig:V612_Sct_main_spec}
\end{center}
\end{figure*}

\noindent \textbf{V612 Sct (ASASSN-17hx):} the very slow Galactic nova V612~Sct was discovered by ASAS-SN on 2017 Jun 19.41 as ASASSN-17hx (see \citealt{ATel_10523} and \citealt{ATel_10524}). The nova was monitored extensively by the ARAS group, and it showed Balmer and He I lines before developing strong Fe II features. The early spectroscopic evolution of the nova and its optical light curve are presented in Figure~\ref{Fig:V612_Sct_main_spec}. The nova rose to peak in 40 days. A long gap in the spectroscopic monitoring of the nova due to solar conjunction prevents us from building a complete spectral evolution. It also prevents us from constraining $t_2$, which we estimate it to be longer than 172 days, implying that V612~Sct is a very slow evolving nova. Nevertheless, the early spectra (days 6 and 10) show that the nova exhibits an early He/N phase before the Fe II P Cygni profiles dominated the spectra (after day 11).\\

\begin{figure*}
\begin{center}

\includegraphics[width=0.62\textwidth]{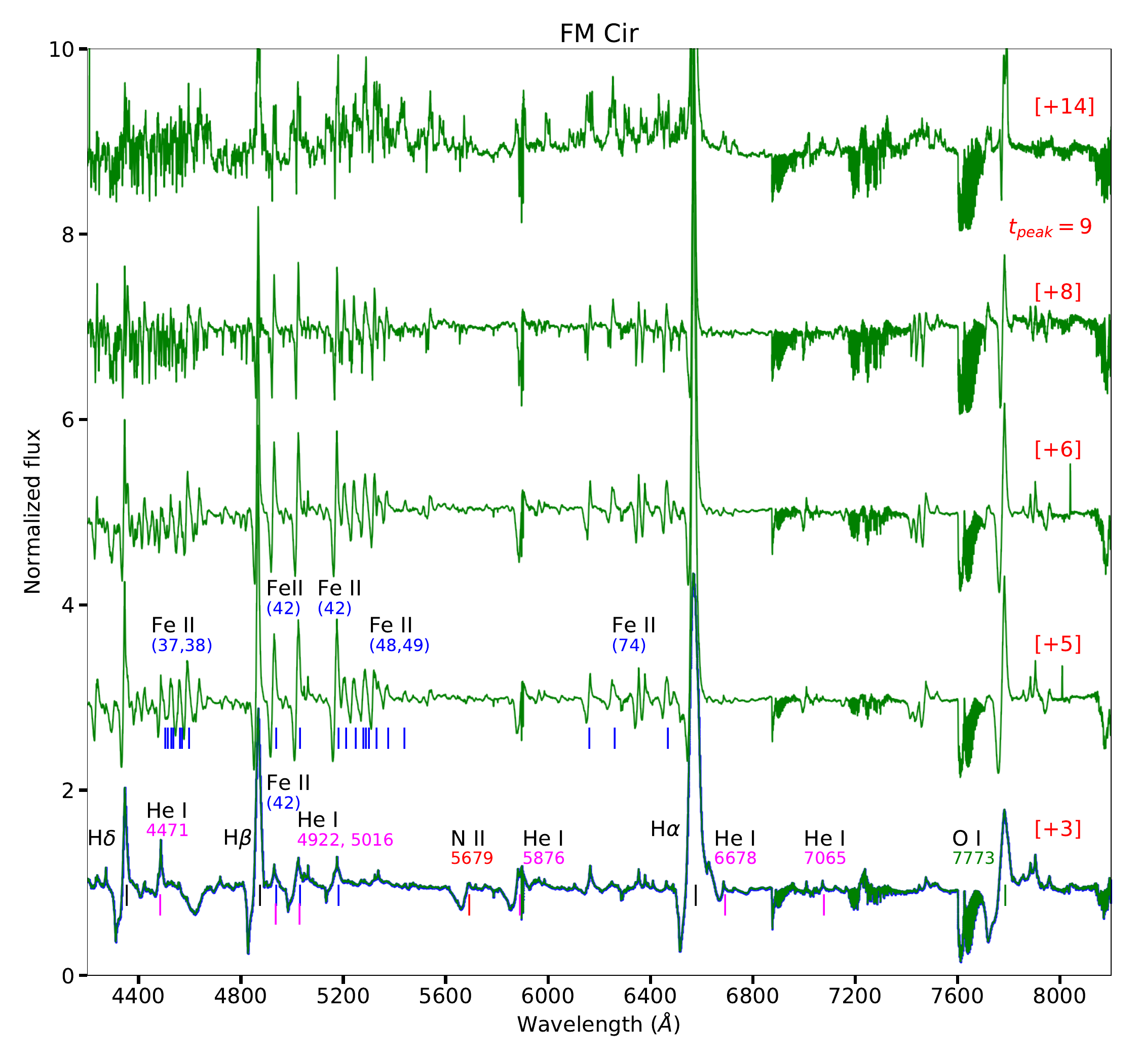}\includegraphics[width=0.388\textwidth]{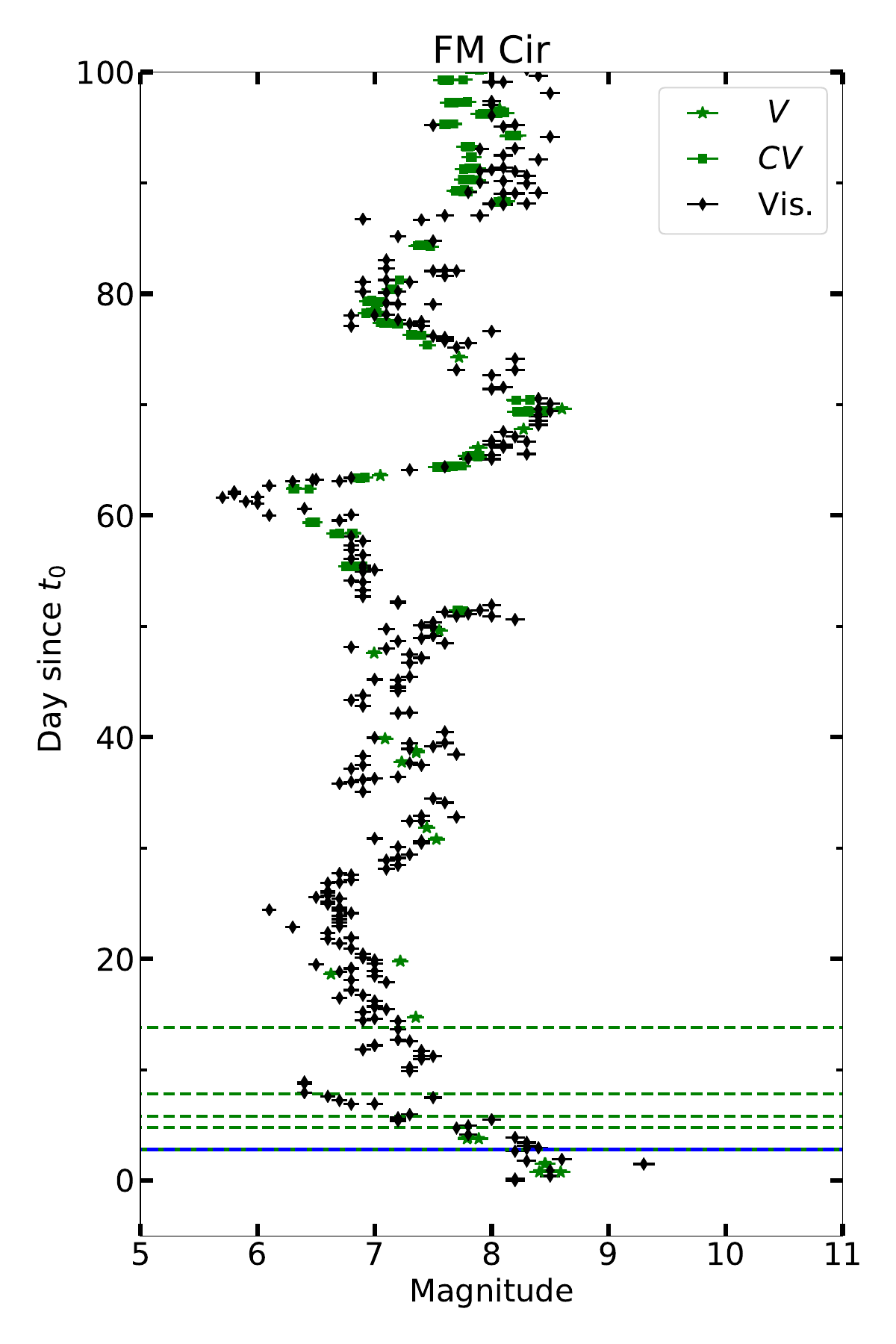}
\caption{Same as Figure~\ref{Fig:T_Pyx_main_spec} but for nova FM~Cir. The spectrum on day 3 is plotted in both green and blue, indicating that this epoch is a transition from phase 1 to phase 2.}
\label{Fig:FM_Cir_main_spec}
\end{center}
\end{figure*}

\noindent \textbf{FM Cir:} the slow Galactic nova FM Cir was discovered by John Seach on 2018 Jan 19.7 and classified as an Fe II nova by \citealt{ATel_11209}. Pre-discovery observations from ASAS-SN show that the eruption started before or by 2018 Jan 17.87. The early spectroscopic evolution of the nova and its optical light curve are presented in Figure~\ref{Fig:FM_Cir_main_spec}. The nova evolution was very slow, with $t_2$ = 120 days, while it took 25 days since $t_0$ to reach optical peak. The first spectroscopic epoch taken on day 3 shows a combination of Balmer and both \eal{Fe}{II} and He/N P Cygni lines, implying that the nova was first observed during a transition phase between phase 1 and phase 2. Thereafter, the spectra are dominated by Balmer and \eal{Fe}{II} P Cygni lines. We only focus on the early spectral evolution during the first two weeks of nova FM~Cir to highlight that traditionally classified \eal{Fe}{II} slow novae also show an early He/N phase, emphasizing that the universal spectral evolution proposed in this work is ubiquitous. We lack further follow up to show the complete spectral evolution of this slow nova.

\subsection{The very fast novae (traditionally He/N)}
In this section we present the early (first few days) spectral evolution of two very fast novae, which are traditionally classified as He/N novae. We show that even very fast novae exhibit a similar spectral evolution to the one described in our main sample. However, the main difference lies in the duration of phases 1 and 2, which tend to be very rapid (a few hours to a couple of days) and thus are easily missed. Nevertheless, dedicated follow up for these two very fast novae during the early hours/days of the eruption provide us with an opportunity to showcase their early spectral evolution. Below we describe the two fast novae:
\\

\begin{figure*}
\begin{center}
  \includegraphics[width=0.62\textwidth]{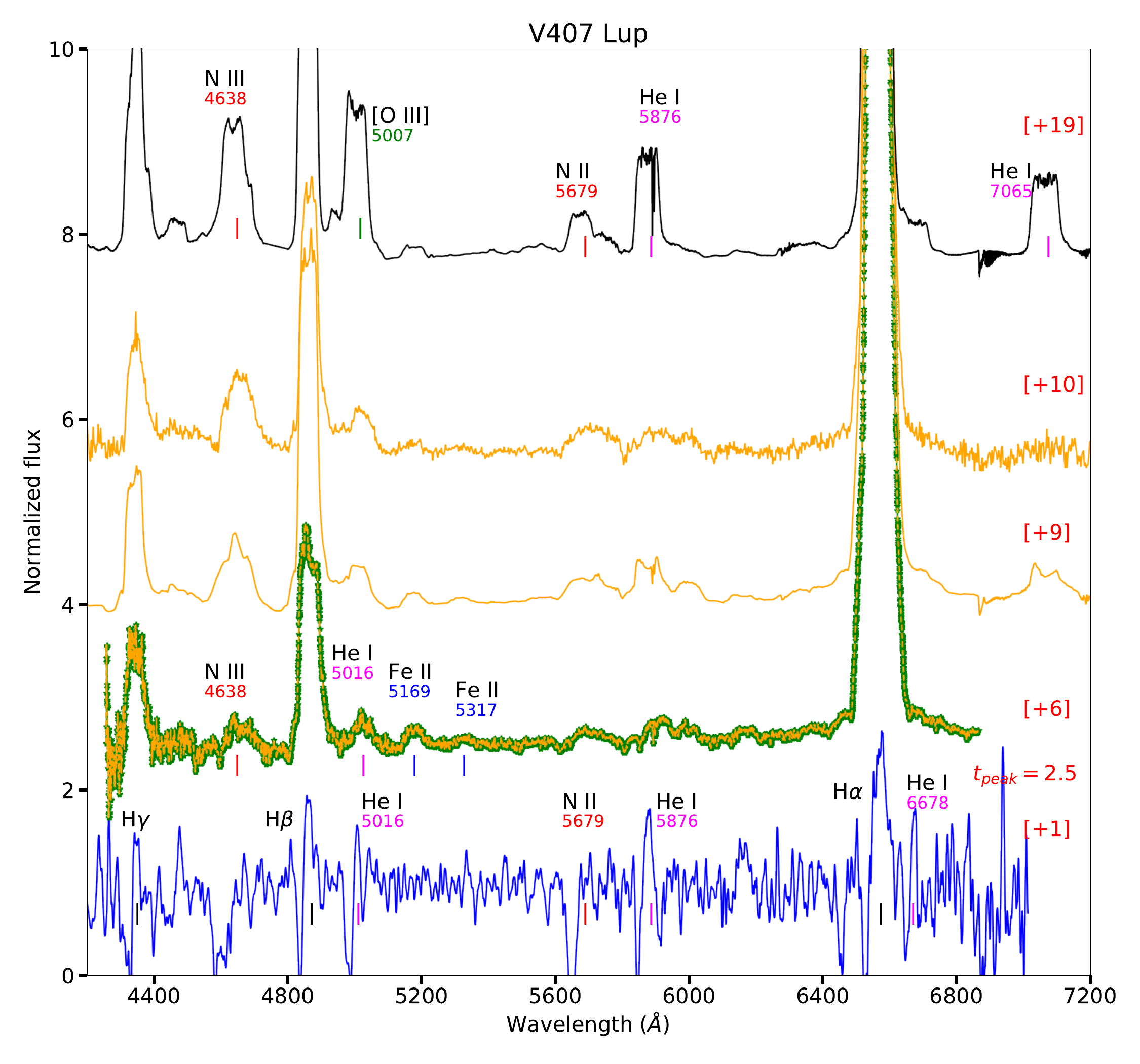}\includegraphics[width=0.388\textwidth]{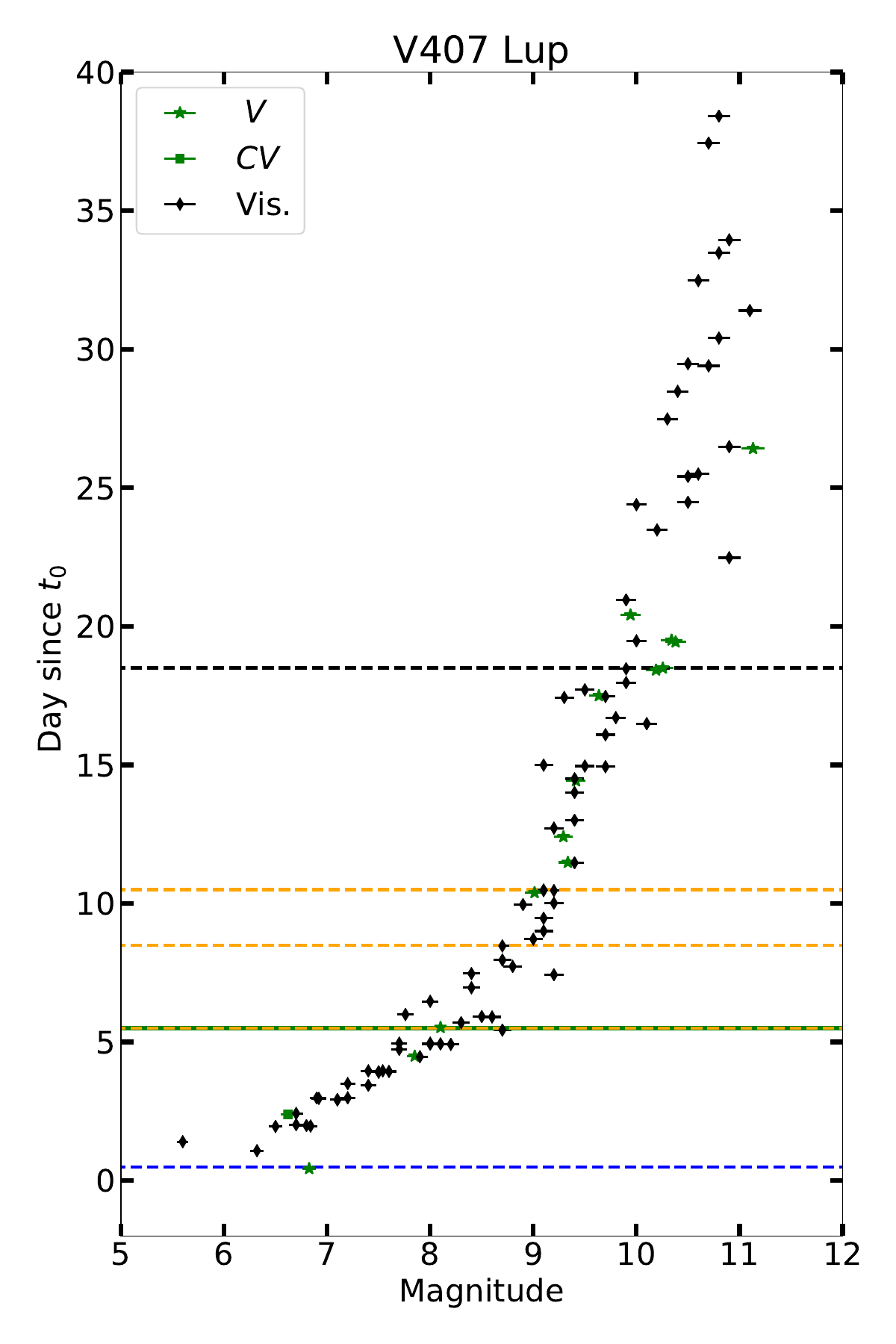}
\caption{Same as Figure~\ref{Fig:T_Pyx_main_spec} but for nova V407~Lup. The spectrum on day 6 is plotted in both green and red, indicating that this epoch is a transition from phase 2 to phase 3.}
\label{Fig:V407_Lup_main_spec}
\end{center}
\end{figure*}

\noindent \textbf{V407 Lup:} the very fast nova V407~Lup was discovered by ASAS-SN as ASASSN-16kt on 2016 Sep 24.00 (HJD 2457655.5 = $t_0$; \citealt{ATel_9538,ATel_9539}). The nova is characterized by $t_2 < $3 days \citep{Aydi_etal_2018_2}, implying that it is one of the fastest novae in the past decade. The nova rose to optical peak from the time of discovery in 1.4 days, highlighting its rapid evolution. The spectroscopic evolution of the nova over the first three weeks of eruption and its optical light curve are presented in Figure~\ref{Fig:V407_Lup_main_spec}. The first spectroscopic epoch, obtained less than a day after discovery, shows P Cygni lines of Balmer and He I. The second epoch was already 5 days later, showing significant differences in comparison to the earlier epoch. The spectrum is now dominated by broad emission lines of Balmer, He I, and N III, with some weak Fe II emission lines, at 4924, 5018, and 5169\,\AA\,(multiplet 42) and at 5317\,\AA\,(multiplet 48). The lines at 4924 and 5018\,\AA\,are likely blended and dominated by nearby He I lines at 4922 and 5016\,\AA\,(see Figure~\ref{Fig:V407_Lup_spec_zoom}). Given the line blending in this particular region, it is challenging to confirm the identity of all the lines in the spectra. The Fe II lines at 5169 and 5317\,\AA\,could conceivably be N II lines 5176\,\AA\,(multiplet 70) and 5320\,\AA\,(multiplet 69), which has been suggested to be present in the spectra of some other novae (e.g., \citealt{Munari_eatl_2006}). However, N II 5320\,\AA\,(69) is a quintet transition where the upper level is  an autoionization state with a large excitation potential $\approx$ 30\, eV \citep{1975aelg.book.....B}. Because N III has a doublet 2p\,$\mathrm{{}^2P^o}$ ground state, a quintet multiplet like N II (69) cannot be directly populated by either electron recombination with N III or by resonance scattering of N II with its 2p$^2$\,$\mathrm{{}^3P}$ ground state. If N II 5320\,\AA\,of the (69) multiplet were to be present, another line of the same multiplet should be present at 5351\,\AA\,with similar intensity \citep{Moore_1945}, which is not the case as shown in Figure~\ref{Fig:V407_Lup_spec_zoom}. Moreover, the Fe II lines multiplet (42) and (48) are more common in nova spectra and are easier to form, compared to the N II lines multiplet (69) and (70). All this makes the identification of these lines as Fe II (42) and (48), rather than N II (69) and (70), more likely. 

Due to the presence of these weak Fe II emission lines in the spectra of V407~Lup, \citet{Izzo_etal_2018} suggested that the nova had a very rapid iron curtain phase (see Section~\ref{sec4.1}). We also suggest that indeed, during this epoch the nova was captured in a transition phase between a rapid phase 2 and phase 3. It is possible that if the nova was observed between day 1 and day 6, we might have detected stronger Fe II lines, however, no observations are available during this phase. The He/N emission lines remained strong until day 10. By day 19, strong forbidden lines of O and N emerged, indicating that the nova has approached/entered the nebular phase. The spectroscopic evolution of nova V407~Lup again shows, that very fast novae, which are traditionally classified as He/N novae, also goes through the 3 main phases proposed in this work, while the first 2 phases (early He/N and Fe II) are very rapid compared to other novae. Moreover, phase 2 here is not as pronounced as for the novae in our main sample, since only a few Fe II lines emission lines are marginally detected. 
\\

\begin{figure*}
\begin{center}
  \includegraphics[width=0.62\textwidth]{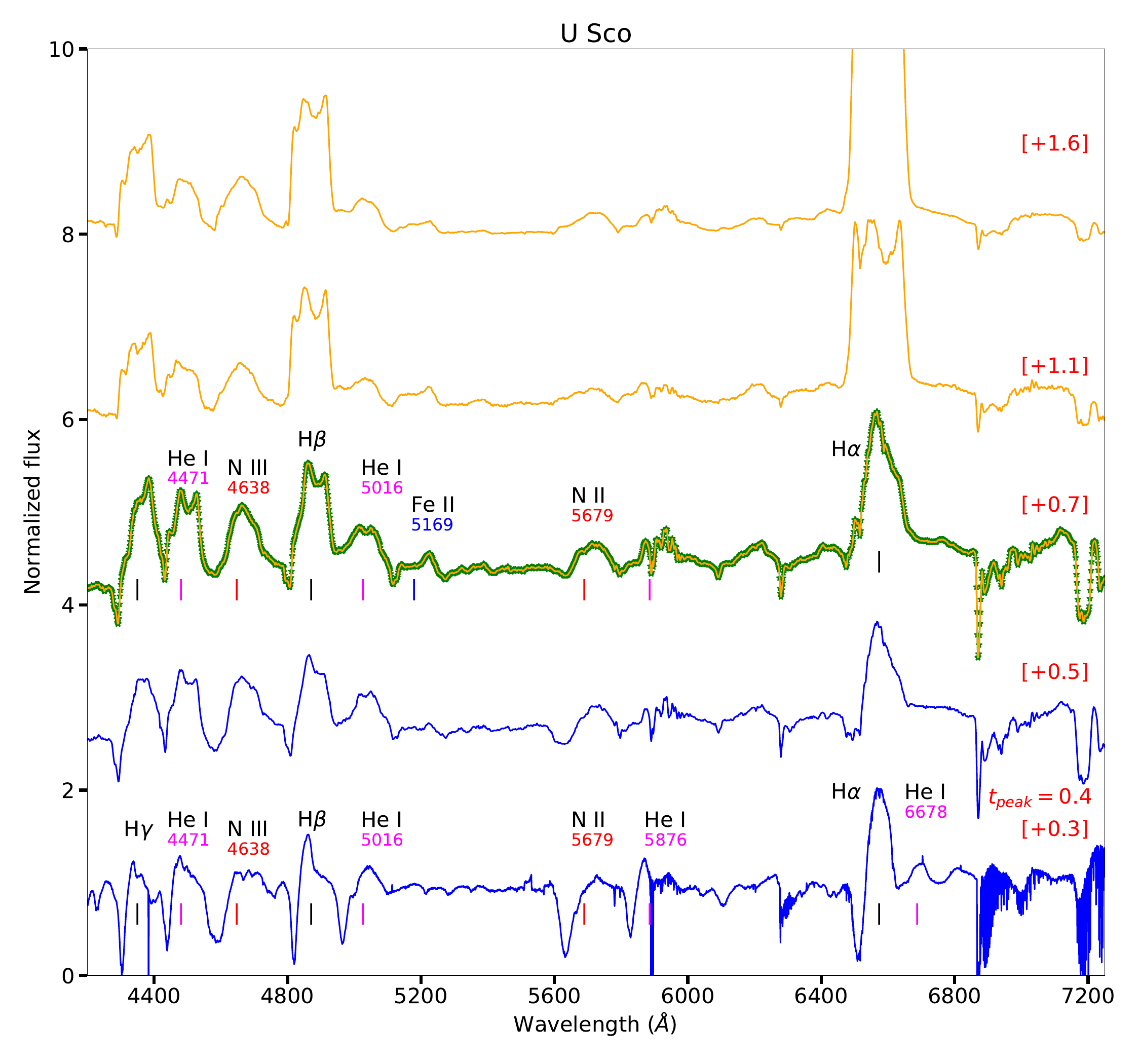}\includegraphics[width=0.388\textwidth]{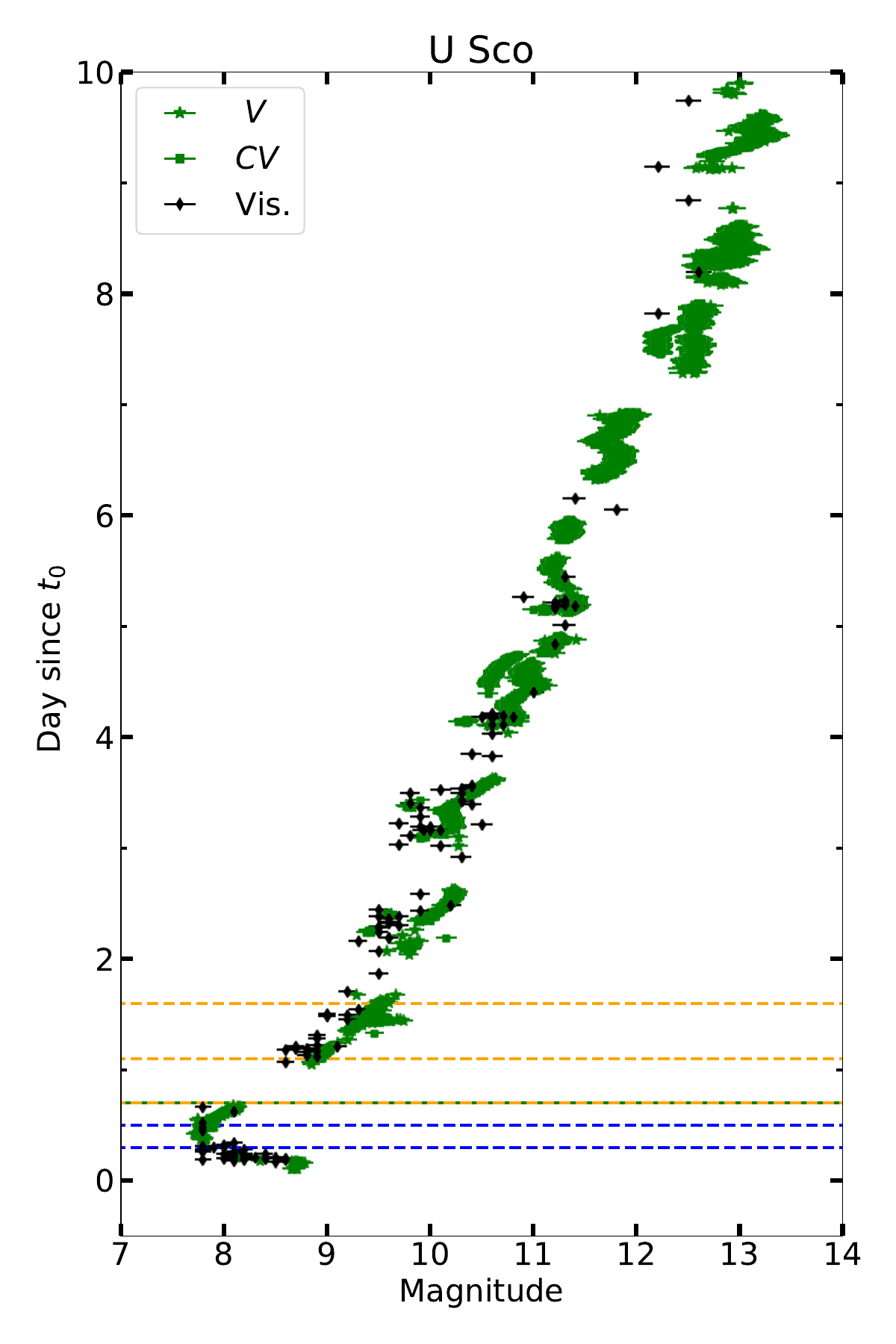}
\caption{Same as Figure~\ref{Fig:T_Pyx_main_spec} but for nova U~Sco. The spectrum on day 0.7 is plotted in both green and red, indicating that this epoch is a transition from phase 2 to phase 3.}
\label{Fig:U_Sco_main_spec}
\end{center}
\end{figure*}

\noindent \textbf{U~Sco (2022):} the 2022 eruption of the recurrent Galactic nova U~Sco was discovered on 2022-06-06.72 (HJD 2459737.22 = $t_0$) by Masayuki Moriyama. The eruption is characterized by a very rapid decline with $t_2 \approx 2 \pm 0.5$ days. It also rose to visible peak from $t_0$ in just half a day. The early spectroscopic evolution of the nova and its optical light curve are presented in Figure~\ref{Fig:U_Sco_main_spec}. Nova U~Sco shows extremely rapid changes in the spectra during the early days of the eruption, however, due to unprecedented dedicated follow up by both professional and citizen scientists, it was possible to disentangle the spectroscopic evolution of the nova during the first 2 days of the eruption with an immaculate cadence. The first spectroscopic epoch, taken 0.3 days after discovery, showed P Cygni lines of Balmer, He I, and N III. By day 0.7, the emission lines strengthened relative to the absorptions, and relatively weak Fe II lines of the (42) multiplet emerged. This possibly indicates that the nova went through an extremely rapid iron curtain phase, similar to the case of nova V407~Lup (likely even faster). The identification of the Fe II lines is also challenging due to the weakness of these lines and to blending with neighbouring lines. In Figure~\ref{Fig:U_Sco_spec_zoom} we present zoom-in spectral plots focusing on the region of the Fe II (42), (48), and (55) multiplets. The plots show that some Fe II lines of these multiplets are possibly present in the spectra, but are weak relative to other lines of He I and Balmer. Some of them are also possibly blended with nearby stronger He I lines. While some of these lines could be N II 5176\,\AA\,and 5318\,\AA\,of the (70) and (69) multiplets as discussed for nova V407~Lup, the identification of these lines as Fe II (42 and 48 multiplets) is more reasonable, given the arguments presented in the previous section. By day 1.1, the absorption features almost disappear and the spectrum is dominated by broad emission lines of Balmer, He I, N II, and N III. This is an extreme case where spectral phases 1 and 2 lasted less than a day. Similar to V407~Lup, phase 2 here is not very pronounced, given the marginal detection of Fe II lines of the (42) and possibly (48) multiplet. However, the potential identification of some lines as Fe II, argue that very fast novae, like U~Sco and V407~Lup go through a very rapid Fe II phase near peak, as they transition from phase 1 to phase 3.

\section{Discussion}
\label{Disc}
\subsection{A universal spectral evolution}
\label{sec4.1}
The spectral evolution, traditionally dubbed ``hybrid'' (where novae show a transition(s) between He/N and Fe II spectral features or show both He/N and Fe II features simultaneously \citealt{Williams_2012}), has been reported in the literature for several nova (e.g., V5558~Sgr; \citealt{Tanaka_etal_2011} and T~Pyx; \citealt{Shore_etal_2011,Ederoclite_2014,Surina_etal_2014,Arai_etal_2015}), and has been discussed theoretically (e.g., \citealt{Shore_2012,Mason_etal_2018,Shore_2014,Hachisu_Kato_2022}). While the sample was still small, 
\citet{Williams_2012} suggested that all novae might be hybrid, but this needed to be confirmed by observations. \citet{Shore_2012,Shore_2014} suggested that these classes are stages in the evolution of the nova and as the opacity and ionization of the ejecta changes throughout the eruption, the line dominating the spectra change. Therefore, the spectra naturally evolve from one class to another as the ejecta conditions change. Our results presented above are in agreement with these previous suggestions, and indicate that the spectral classes of He/N and Fe II are phases in the evolution of a nova and that the ``hybrid'' (early $\mathrm{He/N\,\,} \longrightarrow \mathrm{Fe~II\,\,} \longrightarrow$ late $\mathrm{He/N\,\,}$) evolution is likely universal in novae, regardless of their speed class. Below we offer a brief descriptive interpretation of the observations and the respective spectroscopic phases:

\begin{itemize}
    \item Phase 1 (early He/N): this is an early phase of the eruption during the early rise to optical peak, at a stage where the nova ejecta are still hot (A few times 10$^{4}$\,K; \citealt{Hauschildt_2008}) and dense (of the order of 10$^{10}$\,cm$^{-3}$; \citealt{Metzger_etal_2016}). That is, the presence of high-excitation lines of He I, N II, and N III, with P Cygni profiles implies that the nova ejecta are optically thick and relatively hot, with  photospheric temperature of the order of a few times 10$^4$\,K \citep{Hauschildt_2008,Hachisu_Kato_2022}. The high density is mostly due to the fact that the ejecta  have not yet had enough time to expand. This early phase is relatively fast, lasting for only a few hours (e.g., nova U Sco) up to a few days (e.g., novae Gaia22alz and T~Pyx). Therefore, it is often missed for most novae. 
    
    \item Phase 2 (Fe II): this phase manifests during the late rise to peak, at optical peak, and during the early decline. As the nova ejecta expand, they cool down rapidly, with the photospheric temperature reaching $\approx$ 8000\,K around optical peak \citep{Hauschildt_2008,Hachisu_Kato_2022}. The expansion of the ejecta induces a recombination wave, leading to the emergence of P Cygni lines from neutral and lightly ionized heavy elements (such as Fe II, O I, and Na I). During this stage of the eruption, strong ultraviolet bands at 1500\,--\,1800\,\AA\,and 2300\,--\,2600\,\AA\,are reprocessed into less opaque, lower energies leading to the emergence of the Fe II lines in the optical --- in the literature, it is often called the \textit{iron curtain stage}. This phase, induced by expansion and cooling, lasts while the ejecta density is still high enough (a few times 10$^{10}$\,cm$^{-3}$; \citealt{Shore_2008}) to produce the low-ionization Fe II lines, before the density is too reduced by the continuing expansion (for a more detailed review, see \citealt{Shore_2008,Shore_2012,Shore_2014}). Therefore, for very fast novae with rapidly expanding ejecta, this Fe II phase lasts for a short period  (hours to a couple of days), such as the case of novae V407~Lup and U Sco. In this case the rapidly expanding ejecta might be photo-ionized to near nebular conditions in short period of time (hours/a few days), meaning that the \textit{iron curtain stage} might not take place or take place for a very short time, so lines could appear at the same epoch from different regions in the ejecta, characterized by different densities \citep{Shore_2008}. This could explain the marginal detection of weak Fe II lines in the early spectra of U~Sco and V407~Lup. However, for very slow novae with slowly expanding ejecta, the Fe II phase lasts for weeks and even months (e.g., V1405 Cas, FM~Cir, and V612~Sct). Quoting from \citet{Shore_2008}, ``\textit{we can say that virtually all novae pass through some Fe II-like stage}''. 

    \item Phase 3  (late He/N): this phase typically manifests during the late decline (2 to 5 magnitudes below optical peak). As the nova ejecta expand further and the mass-loss rate decreases, the ejecta drop in density and the photosphere recedes backwards to inner hotter regions due to the decrease in the optical depth ($\propto t^{-2}$). This leads to an increase in the ionization of the ejecta and eventually the \textit{lifting of the iron curtain}. In the optical, this manifests as the emergence of emission lines of high-excitation transitions such He I, He II, N II, N III, and [N II] \citep{Hachisu_Kato_2022}. Unlike, phase 1 where similar lines show P Cygni profiles, at this stage the lines show mostly emissions as the P Cygni absorptions become shallower. 
\end{itemize}

The continuously expanding ejecta drop in density even further to a stage where the entire ejecta eventually become transparent and completely ionized, leading to the emergence of auroral and nebular lines particularly single and double ionized forbidden lines of oxygen, nitrogen, and eventually iron \citep{Shore_2008}. This so-called nebular phase succeeds phase 3 in most novae. Nevertheless, \citet{Williams_1992} showed that some novae do not develop strong nebular lines. This is mostly determined by the ejecta conditions (temperature and density) late in the eruption, e.g., if the supersoft emission \citep{Wolf_etal_2013} turns off before the ejecta drop enough in density, it is possible that a nebular phase does not develop in the spectra \citep{Cunningham_etal_2015}.

We suggest that most if not all novae go through the three phases listed above, and that the Fe II and He/N spectroscopic classes are rather evolutionary phases during the eruption. That is, novae which are traditionally classified as Fe II novae are those that are caught during phase 2 near/after optical peak. Meanwhile, novae which are traditionally classified as He/N type are caught in phase 3 near/after peak, and represent systems wherein the first two phases are very rapid (lasting for $\sim$ hours), and are therefore missed. Novae that show features from both classes simultaneously are likely going through a transition between the phases (e.g., spectra of days 7 and 48 for T Pyx in Figure~\ref{Fig:T_Pyx_main_spec}, day 3 spectrum of FM Cir in Figure~\ref{Fig:FM_Cir_main_spec}, and day 6 spectrum of V407 Lup in Figure~\ref{Fig:V407_Lup_main_spec}). The line-driving presumably works primarily on the ion species that are prominent in the observed spectra at the time (Fe in the Fe II phase, He and N during the He/N phase), with other ion species being carried along passively via Coulomb friction. Whether the conditions are right for this to create regions of different abundances (see, e.g., \citealt{Huger_Groote_1999})
should perhaps be investigated in the future.

One might argue that the (He/N $\longrightarrow$ Fe II $\longrightarrow$ He/N) evolution is only common in slowly evolving novae. However, since our sample consists of fast and slow novae, and since we see evidence for rapid phases 1 and 2 in very fast novae like U~Sco, V407~Lup, and V659~Sct, our data indicate that the (He/N $\longrightarrow$ Fe II $\longrightarrow$ He/N) evolution is likely universal. 

\subsection{Origin in different bodies of gas?}
\citet{Williams_2012} suggested that He/N spectra are consistent with an
origin in the white dwarf (WD) ejecta, whereas Fe II spectra point to formation in a large circumbinary
envelope of gas whose origin is the secondary star. If this is the case, we should observe a large gradient in velocities between the Fe II lines in comparison to the He/N and Balmer lines, for the same nova --- since the nova ejecta will be characterized by larger velocities (a few thousands km\,s$^{-1}$) compared to circumbinary-medium gas (dozens of km\,s$^{-1}$, e.g., as observed in symbiotic systems \citealt{Mikolajewska_2012,Munari_2019}). However, \citet{Aydi_etal_2020b} showed in their study of the early spectral evolution of novae and the dynamics of their ejecta that the Balmer lines, the Fe II lines (particularly the prominent 42 multiplet), He lines, and the transient heavy elements absorption (THEA) lines all have similar velocities and show similar evolution. This led \citet{Aydi_etal_2020b} to conclude that the Fe lines and transient heavy elements absorptions (THEA) lines all originate in the nova ejecta, rather than originating in circumbinary material. 

\begin{figure}
\begin{center}
  \includegraphics[width=\columnwidth]{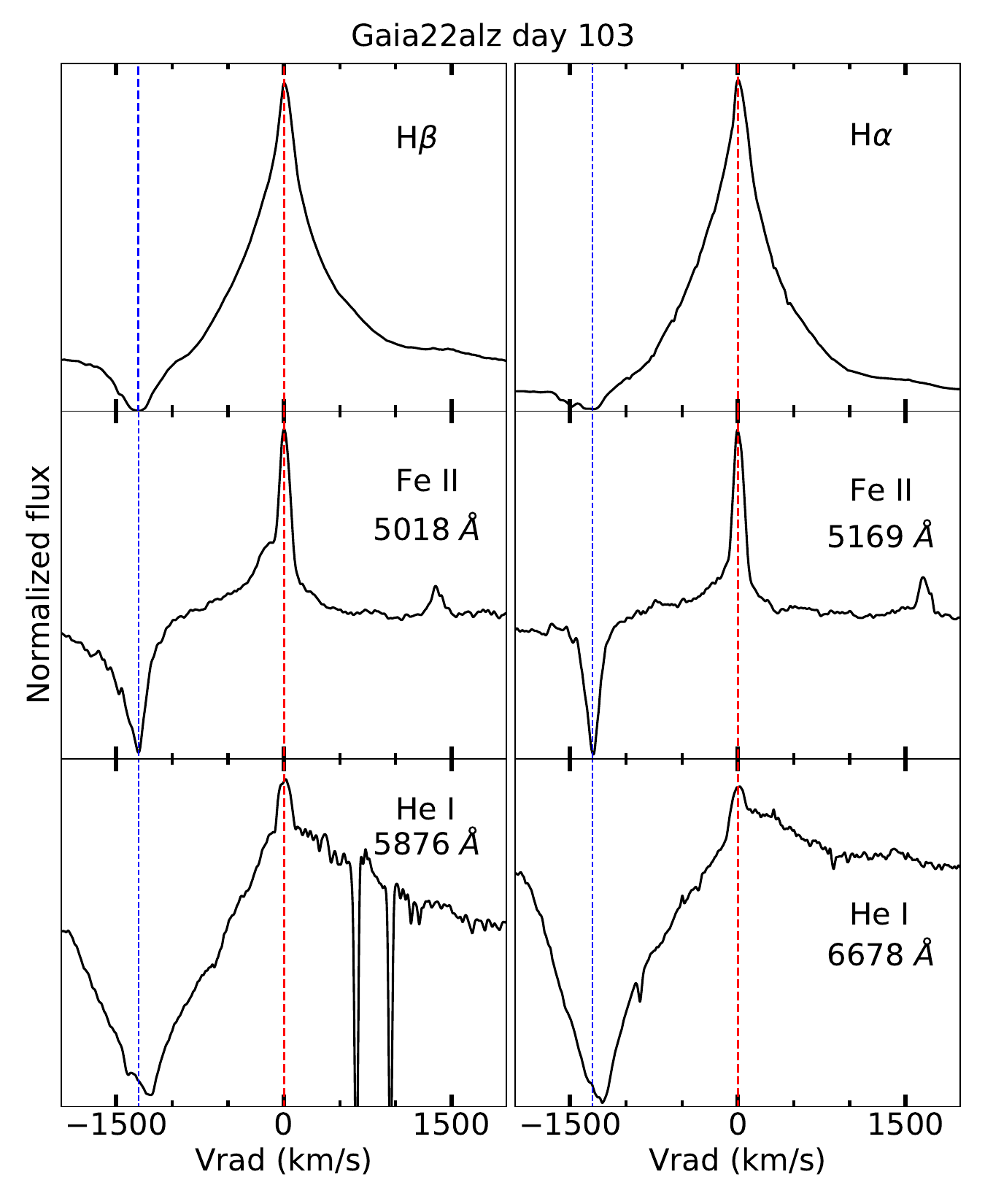}
\caption{The profiles of Balmer (top), Fe II (middle), and He I (bottom) lines for nova Gaia22alz as observed 103 days after $t_0$. The red dashed lines represent the peak wavelength of each line, while the green dashed lines represent $v = -1300$\,km\,s$^{-1}$ relative to the line peak.}
\label{Fig:Gaia22alz_profiles}
\end{center}
\end{figure}

In Figure~\ref{Fig:Gaia22alz_profiles} we show line profiles of Balmer, Fe II, and He I for nova Gaia22alz (one of the best monitored novae in our sample) on day 103, at a stage of transition between phases 1 and 2, when both He and Fe features are present in the spectrum. The absorption troughs of the Balmer, Fe II, and He I P Cygni lines all have comparable blueshifted velocities ranging between 1200 and 1300\,km\,s$^{-1}$, implying that they originate in the same body of gas, that is, the nova ejecta. Nevertheless, the Fe II line structure is slightly different compared to the He I lines, which is not surprising since lines of different ionization potentials might originate at different depths/locations in the ejecta.

\subsection{Interpretation in context of multiple ejections/outflows}

Nova spectra are known to show multiple absorption and emission components at distinct velocities for the same species/lines, ranging between a few hundred up to a few thousand km\,s$^{-1}$ (e.g., \citealt{Mclaughlin_1945,Payne-Gaposchkin_1957}). There have long been suggestions that different line components are associated with different outflows or ejected shells (e.g., \citealt{Friedjung_1966_I,Friedjung_2011,Arai_etal_2016}). More recently, \citet{Aydi_etal_2020b} suggested that all novae show evidence for at least two physically distinct outflows: an initial slow one manifesting as pre-maximum P Cygni profiles characterized by low velocities (a few hundred km\,s$^{-1}$), followed by a faster outflow which manifests as broader P Cygni components and emission lines with velocities of a few thousands km\,s$^{-1}$, emerging after maximum visible brightness. \citep{Shen_Quataert_2022} suggested that the bulk (80--90\%) of the nova ejecta is carried by the slow flow, while the fast flow is less dense and carries a small fraction of the nova ejecta.
So how does the spectral evolution suggested in this work fit with the multiple outflows scenario presented in previous studies such as \citet{Mclaughlin_1945,Friedjung_1987,Friedjung_2011,Arai_etal_2016,Aydi_etal_2020b}? Do certain lines or ionization species originate in one of these outflows only? 

The first He/N phase likely originates in the slow outflow during the early stages of its expansion. This early expansion of the slow flow manifests as P Cygni lines of Balmer and He/N, characterized by slow velocities (a few 100s km\,s$^{-1}$), which is the case for all the novae in our sample (see Figures 1 to 8). As the slow flow further expands, it cools down due to mainly adiabatic expansion, leading to a recombination phase, and the spectrum enters phase 2 (Fe II), developing P Cygni profiles of Balmer and Fe II (see, e.g., Figure~\ref{Fig:T_Pyx_main_spec}). 

After peak, fast components emerge and are easily detected in some of the species/transitions.
While the interpretation of \citet{Aydi_etal_2020b} was mostly based on the evolution of Balmer lines, where features from both outflows are easily visible, they also show that lines from heavier elements such as the Fe II (42) multiplet and some He I lines also show evidence for these two outflows. However, the less dense faster flow might not produce broad emission features in some of the transitions of heavy elements. For example, \citet{Aydi_etal_2020b} showed that the Fe II (42) multiplet lines develop fast components, while the (40) and (46) multiplet lines do not show such fast components. They suggest that the absence of fast components in these multiplets is due to their lower oscillator strength values, which might result in weak broad emission components from the low-density fast flow which are hardly detected. In slow novae, the ejecta is much denser compared to very fast novae, and therefore, even the fast flows have densities high enough to produce fast components in some of the  Fe II lines, such as multiplet (42). However, in fast novae, the low density fast flows might not develop any Fe II features.

As the outflows collide and merge, an intermediate velocity component emerge, which was named the principal component by \citet{Mclaughlin_1945} and is suggested to become the dominant component in the spectra \citep{Friedjung_1987}. The velocity of the principal component is intermediate between the slow and fast components. The slow flow disappears shortly after peak, while the broad emission lines from the fast component become more difficult to resolve, as their intensity drops in comparison to the principal component which then dominates (see \citealt{Aydi_etal_2020b}). As the eruption evolves, lines of He I, N II, and, [N II] develop, characterized by velocities equivalent to that of the principal component. This evolution could either occur over weeks for slowly evolving novae or over hours/days for rapidly evolving novae. 

\begin{figure}
\begin{center}
  \includegraphics[width=\columnwidth]{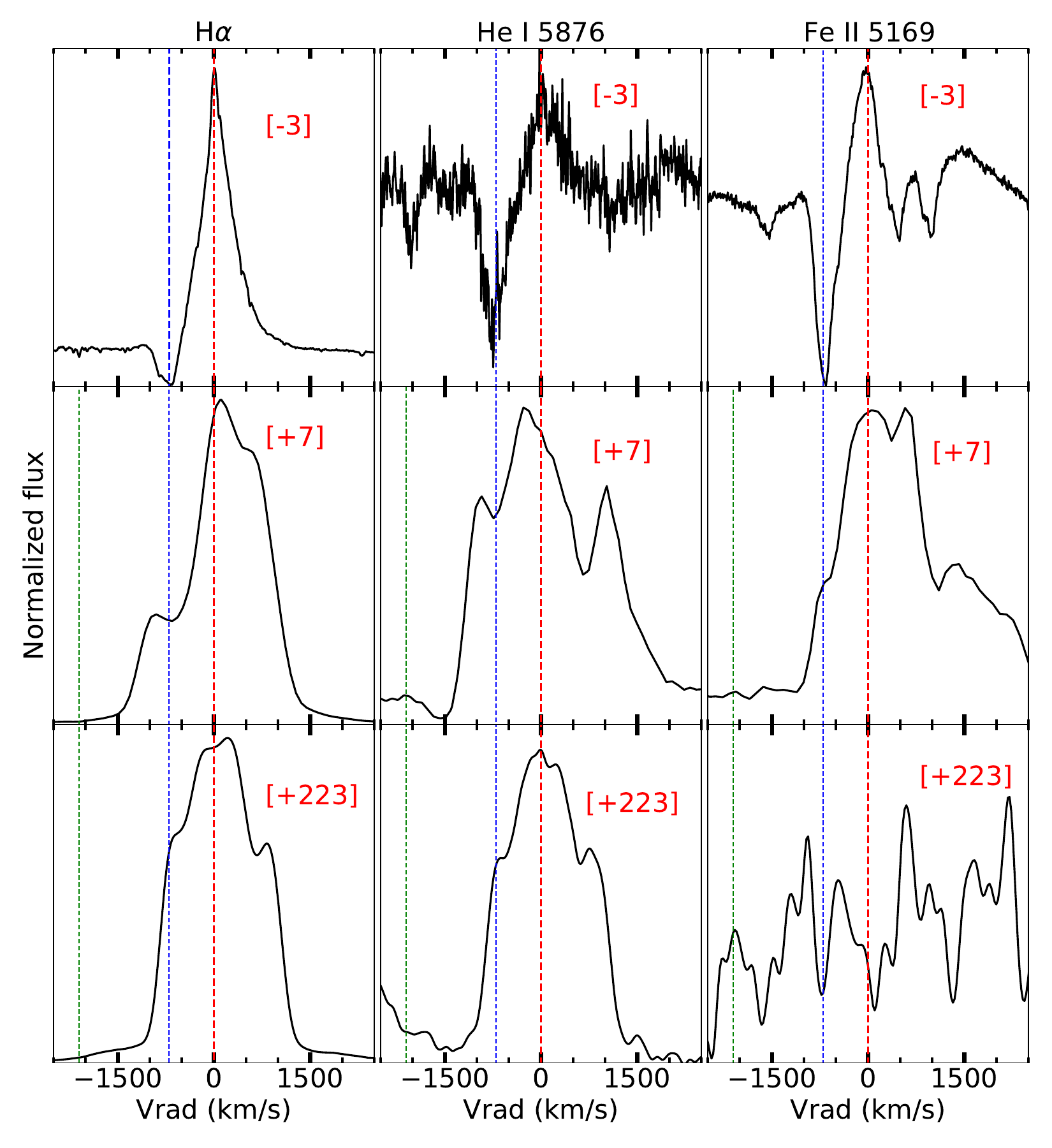}
\caption{The profiles of H$\alpha$ (left), He I (middle), and Fe II (right) lines for nova V1405~Cas. The numbers between brackets are days relative to visible peak brightness (day 56). The red dashed lines represent the central wavelength of each line, while the blue and green dashed lines represent $v = -700$\,km\,s$^{-1}$ and $v = -2100$\,km\,s$^{-1}$ relative to line center, respectively.}
\label{Fig:V1405_Cas_line_profiles}
\end{center}
\end{figure}

As an example, we plot in Figure~\ref{Fig:V1405_Cas_line_profiles} the line profiles of H$\alpha$, Fe II 5169\,\AA, and He I 6678\,\AA\,for nova V1405~Cas, one of our best observed novae. We select these lines because they are not blended with nearby lines. A few days before peak (during phase 2), we see P Cygni profiles with absorption troughs at velocities of around $-700$\,km\,s$^{-1}$. During this phase, the He I lines are much weaker compared to the Fe II lines, but are still detectable. The 700\,km\,s$^{-1}$ reflects the velocity of the slow flow. After peak brightness, H$\alpha$ shows a faster component reaching 2100\,km\,s$^{-1}$. This velocity reflects the velocity of the fast flow. The Fe II and He I lines show velocities between 1000 and 1500\,km\,s$^{-1}$, likely reflecting that of the intermediate (principal) component \citep{Friedjung_1987}. As the nova reaches phase 3, the Fe II lines disappear, while the He I emission lines show velocities of around 1200\,km\,s$^{-1}$, again reflecting the velocity of the intermediate (principal) component.

\subsection{Interpretation in context of multi-wavelength emission}
Novae are panchromatic transients, emitting from radio to $\gamma$-rays (e.g., \citealt{Chomiuk_etal_2021}). As the nova ejecta expand, the peak of the spectral energy distribution shifts from optical to higher energies (first UV and then supersoft X-rays; \citealt{Gallagher_Starrfield_1976,Gallaher_etal_1978,Page_etal_2013}). More recently, the GeV $\gamma$-ray detection by the Large Area Telescope on board of the Fermi $\gamma$-ray satellite, highlighted the presence of strong energetic shocks \citep{Ackermann_etal_2014,Cheung_etal_2016,Franckowiak_etal_2018}, which might contribute a significant fraction of the nova visible luminosity \citep{Metzger_etal_2015,Li_etal_2017_nature,Aydi_etal_2020a}. These shocks could also be the source of non-thermal hard X-ray (2 to 80\,keV; \citealt{Mukai_Ishida_2001,Mukai_etal_2008,Aydi_etal_2020a,Sokolovsky_etal_2020,Gordon_etal_2021,Sokolovsky_etal_2022,Sokolovsky_etal_2023}) and synchrotron radio emission \citep{Finzell_etal_2018,Chomiuk_etal_2020}. The shocks are suggested to occur at the interface of multiple phases of mass-loss, more particularly between an initial slow outflow, followed by a faster one (e.g., \citealt{Chomiuk_etal_2014,Metzger_etal_2015,Aydi_etal_2020a,Aydi_etal_2020b}). How does the optical spectral evolution presented in this work fit with the panchromatic emission in novae?

During the early rise to peak (early He/N phase), the ejecta are optically thick, and the emission from the remnant burning on the surface of the hot white dwarf is mostly absorbed by the dense ejecta, where it escapes in the visible. This is still true during the late rise to peak, when the spectra are cooling down, but still dense, exhibiting a shift into an Fe II spectra. Near peak, faster components emerge in different line species, coinciding with the detection of GeV $\gamma$-rays, originating from shock interaction (see previous section). Hard (2\,--\,80\,keV) X-ray emission from the shocks are also detected \citep{Sokolovsky_etal_2020,Sokolovsky_etal_2022}, especially as the ejecta expand further and become more transparent \citep{Gordon_etal_2021}. Similarly, non-thermal synchrotron radio emission is observed at some stage during the eruption when the ejecta are transparent to the radio emission \citep{Chomiuk_etal_2020}. This happens during the decline from peak, when the spectra are transitioning from the Fe II stage to the later He/N stage. As the ejecta expand even further (dropping more in density), the spectra start transitioning from the late He/N phase to the nebular phase, coinciding with the detection of supersoft X-ray emission in several novae (e.g., \citealt{Page_etal_2013,Page_etal_2015,Page_etal_2020}). The beginning of the nebular phase usually marks the start of the supersoft emission phase. Around this stage, many novae also show a hike in the thermal radio flux, as the ejecta are now extended enough to produce strong thermal radio emission \citep{Hjellming_eta_1979,Chomiuk_etal_2020}.

\subsection{An early phase preceding phase 1?}
\citet{Arai_etal_2015} reported relatively narrow (FWHM $\approx$ 600\,km\,s$^{-1}$) emission lines of He II, N III, Balmer, and He I in the early spectra of T~Pyx taken 0.2 days after the discovery of its 2012 eruption. These high-excitation lines of He II and N III disappear after a day, and the Balmer and He I lines develop P Cygni absorption components. Similarly, \citet{Aydi_etal_2023a} recently reported high excitation narrow (FWHM $\approx$ 300\,km\,s$^{-1}$) emission lines of N III, He II, and C IV in the early spectra of Gaia22alz, along with Balmer and He I lines. These high-excitation narrow emission lines also disappear as the nova evolves, and the Balmer and He I lines develop P Cygni profiles (see Figure~\ref{Fig:Gaia22alz_main_spec}). The main difference between this early phase and phase 1 is the absence of P Cygni profiles, the presence of high-excitation lines, and the relatively slow velocities (FWHM of the lines if around 300-400\,km\,s$^{-1}$). \citet{Aydi_etal_2023a} discussed different possible origins for these high-excitation narrow emission lines, and they concluded that they likely originate in a hot, optically thin nova envelope, which is not yet fully ejected and is reprocessing some of the high-energy emission from the white dwarf surface during the early UV/X-ray flash phase (e.g., \citealt{Hillman_etal_2014,Konig_etal_2022}).

It might be that this early and rarely observed phase precedes phase 1 in most novae, but it is extremely rapid and often missed. In the era of new all-sky surveys, we expect to detect more novae during their early rise, which would allow spectroscopic follow-up during these early and critical phases. Such early follow up could show spectra consistent with the early spectra of novae T~Pyx and Gaia22alz, proving that this earlier phase, characterized by relatively narrow high-excitation emission lines, is common. 

\subsection{Oscillatory spectral behavior in flaring novae}

The spectroscopic evolution of the flaring nova V1405~Cas (Figure~\ref{Fig:V1405_Sct_main_spec_1}) shows an oscillatory behavior where the spectra oscillate between phases 2 and 3 multiple times, as the nova shows multiple maxima in its optical light curve (Figure~\ref{Fig:V1405_Sct_main_spec_1}). During this stage, the spectra show both Fe II and He/N P Cygni lines, but the relative strength of these features vary. Such a behavior has also been reported previously for other flaring novae (e.g., \citealt{Tanaka_etal_2011,Tanaka_etal_2011_159,Aydi_etal_2019_I}). While determining the origin of the multiple flares/maxima in the optical light curves of novae is outside of the scope of this work, we briefly discuss potential explanations for the oscillatory spectral behavior observed in such novae: (1) a change in the brightness of the central source leading to flares in the optical light curve and eventually a change in the radius of the photosphere. This change in the photosphere radius means that the emission could originate in different regions of the ejecta as a function of time (outer colder regions are dominated by low-ionization Fe II lines, while inner hotter regions lead to higher excitation lines of He I, He II, and N II); (2) multiple ejections, interacting with the previously existing ejecta and leading to flares in the optical light curves. These new ejections could lead to an increase in the density of the ejecta resulting in spectra dominated by Fe II lines. As the ejecta eventually expand, they drop in density leading to the emergence of He/N features. If the nova exhibits multiple ejections, we might observe multiple peaks in the light curve and an oscillatory behaviour in the spectra. \citet{Steinberg_etal_2020} discussed how multiple collisions between alternative fast/slow outflows can produce accelerating and decelerating line features and rapidly varying density structures.

\section{Conclusions}
\label{sec_conc}
In this work, we revisit pioneering studies from the past 2--3 decades \citep{Williams_1992,Williams_2012,Shore_2012,Shore_2013,Shore_2014}, with the aim of discussing the origin of the Fe II and He/N spectroscopic classes of novae and testing some previous suggestions about the universality of certain spectral evolution where novae show features from both of these classes as a function of time since eruption \citep{Shore_2008,Shore_etal_2011,Tanaka_etal_2011,Surina_etal_2014,Ederoclite_2014}. In order to do this, we present detailed spectral evolution of a sample of novae, throughout different phases of their eruptions (early rise, optical peak, early decline, and late decline), which is made possible by the era of new all-sky surveys and the collaboration between professional astronomers and citizen scientists Our data show that novae of different speed classes show a similar spectral evolution going through at least three spectral phases: phase 1 (early He/N) when the spectra are dominated by P Cygni lines of Balmer, He I, N II, and N III; phase 2 (Fe II), when the spectra are dominated by P Cygni lines of Balmer, Fe II, and O I; phase 3 (late He/N), when the spectra are dominated by emission lines of Balmer, He I, He II, N II, and NII. A fourth phase, dubbed the nebular phase, develops in many novae, when nebular forbidden lines of O I, O II, O III, and N II emerge and dominate the spectra. This evolution is primarily driven by changes in the opacity, ionization, temperature, and density of the nova ejecta, as previously suggested by \citet{Shore_2008,Shore_2012,Shore_2014}. 

Based on our dataset, this evolution appears to be ubiquitous, whether the nova evolves rapidly or slowly, emphasizing that the traditional classes of Fe II and He/N are evolutionary phases that most if not all novae go through during their eruptions. The main difference is the duration of these phases, which could last for days/weeks for slow novae or hours/days for very fast novae. Traditionally classified Fe II novae are typically first caught in phase 2 near peak and they tend to be slow novae, where phase 2 lasts for days/weeks. While traditionally classified He/N novae are those typically first observed in phase 3 near peak and they tend to evolve rapidly, where phases 1 and 2 only last for a few hours/days, that is these early phases are easily missed. Therefore, cautious is required when assigning Fe II or He/N spectral classes to novae, as these classes are stages in the spectral evolution of a nova.

\section*{Acknowledgments}

We thank the ARAS and AAVSO observers from around the world who contributed their spectra to the ARAS database and their magnitude measurements to the AAVSO International Database, used in this work.

E.A. acknowledges support by NASA through the NASA Hubble Fellowship grant HST-HF2-51501.001-A awarded by the Space Telescope Science Institute, which is operated by the Association of Universities for Research in Astronomy, Inc., for NASA, under contract NAS5-26555. LC acknowledges NSF awards AST-1751874 and AST-2107070 and a Cottrell fellowship of the Research Corporation. JS was supported by the Packard Foundation. DAHB gratefully acknowledges the receipt of research grants from the National Research Foundation (NRF) of South Africa. AK acknowledges the Ministry of Science and Higher Education of the Russian Federation grant 075-15-2022-262 (13.MNPMU.21.0003).  BDM is supported in part by NASA (grants 80NSSC22K0807 and 80NSSC22K1573).
JM was supported by the National Science Centre, Poland, grant OPUS 2017/27/B/ST9/01940.
KJS is supported by NASA through the Astrophysics Theory Program (80NSSC20K0544).
AE acknowledge the financial support from the Spanish Ministry of Science and Innovation and the European Union - NextGenerationEU through the Recovery and Resilience Facility project ICTS-MRR-2021-03-CEFCA .
A part of this work is based on observations made with the Southern African Large Telescope (SALT), with the Large Science Programme on transients 2021-2-LSP-001 (PI: DAHB). Polish participation in SALT is funded by grant No. MEiN 2021/WK/01. This paper was partially based on observations obtained at the Southern Astrophysical Research (SOAR) telescope, which is a joint project of the Minist\'{e}rio da Ci\^{e}ncia, Tecnologia e Inova\c{c}\~{o}es (MCTI/LNA) do Brasil, the US National Science Foundation's NOIRLab, the University of North Carolina at Chapel Hill (UNC), and Michigan State University (MSU).
Analysis made significant use of \textsc{python} 3.7.4, and the associated packages \textsc{numpy}, \textsc{matplotlib}, \textsc{seaborn}, \textsc{scipy}. 
Data reduction made significant use of \textsc{MIDAS FEROS} \citep{Stahl_etal_1999}, echelle \citep{Ballester_1992}, PySALT \citep{Crawford_etal_2010},  and IRAF \citep{Tody_1986,Tody_1993}.

\section*{Data availability}
The data are available as online material and can be found here: \url{https://www.dropbox.com/scl/fi/3kdnmu52lp8j6j2y7nclp/FeII_HeN_onine_data.zip?rlkey=u8gxya6q970cw18f46k804mqr&dl=0}.\\ Supplementary plots can be found here: \url{https://www.dropbox.com/scl/fi/0glj9qxgj6zoj884qox1k/FeII_HeN_online_plots.zip?rlkey=ml04qp2v2b6ylos5fn97zwvt6&dl=0}.

\bibliographystyle{mnras_vanHack}
\bibliography{biblio}

\appendix

\renewcommand\thetable{\thesection.\arabic{table}}    
\renewcommand\thefigure{\thesection.\arabic{figure}}   
\setcounter{figure}{0}

\section{Supplementary plots and tables}
\label{appB}
In this Appendix we present supplementary plots and tables.

\begin{figure*}
\begin{center}
  \includegraphics[width=0.75\textwidth]{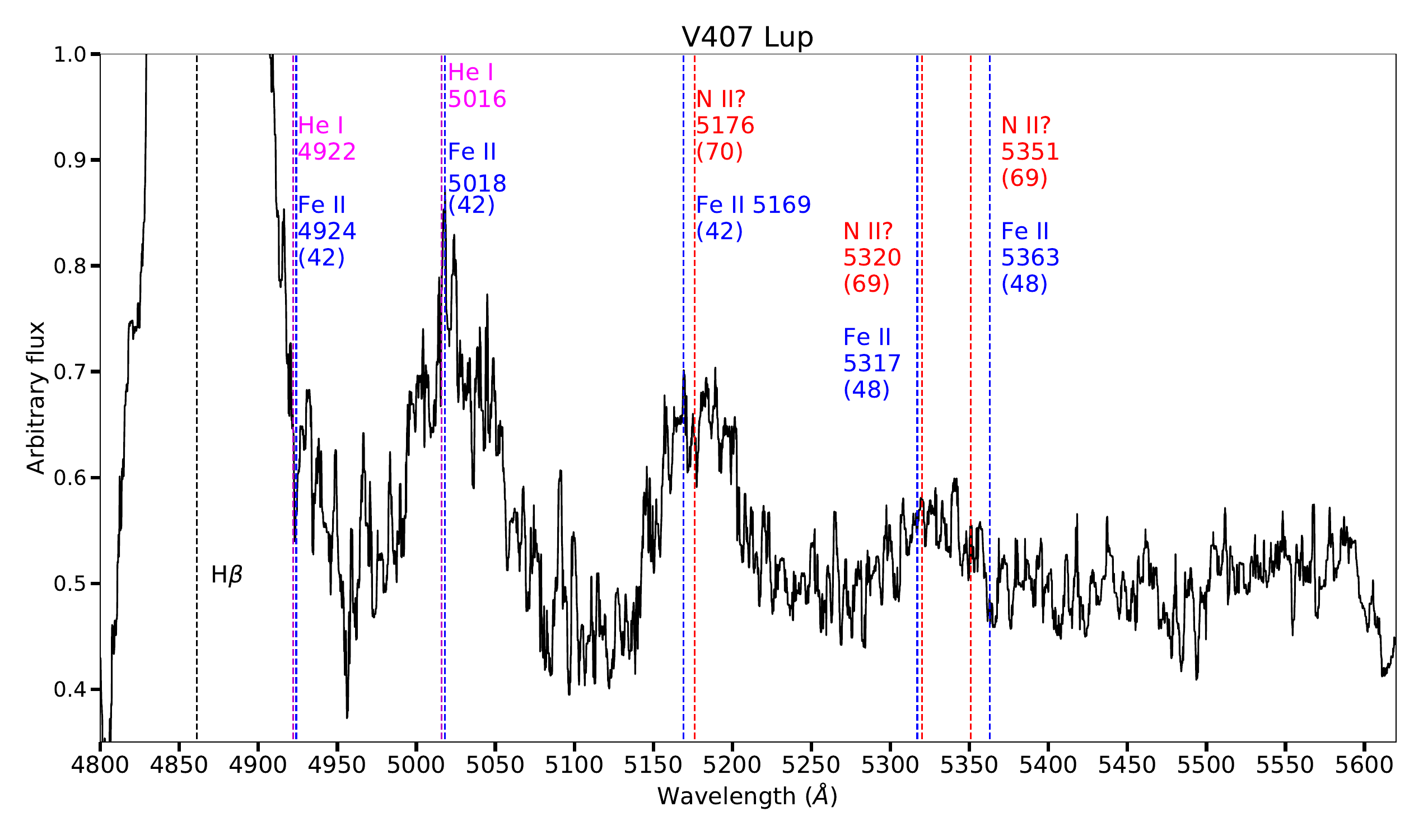}
\caption{A zoom-in spectral plot of nova V407 Lup on day 6, focusing on the region of Fe II (42) and (48) multiplets.}
\label{Fig:V407_Lup_spec_zoom}
\end{center}
\end{figure*}

\begin{figure*}
\begin{center}
  \includegraphics[width=0.75\textwidth]{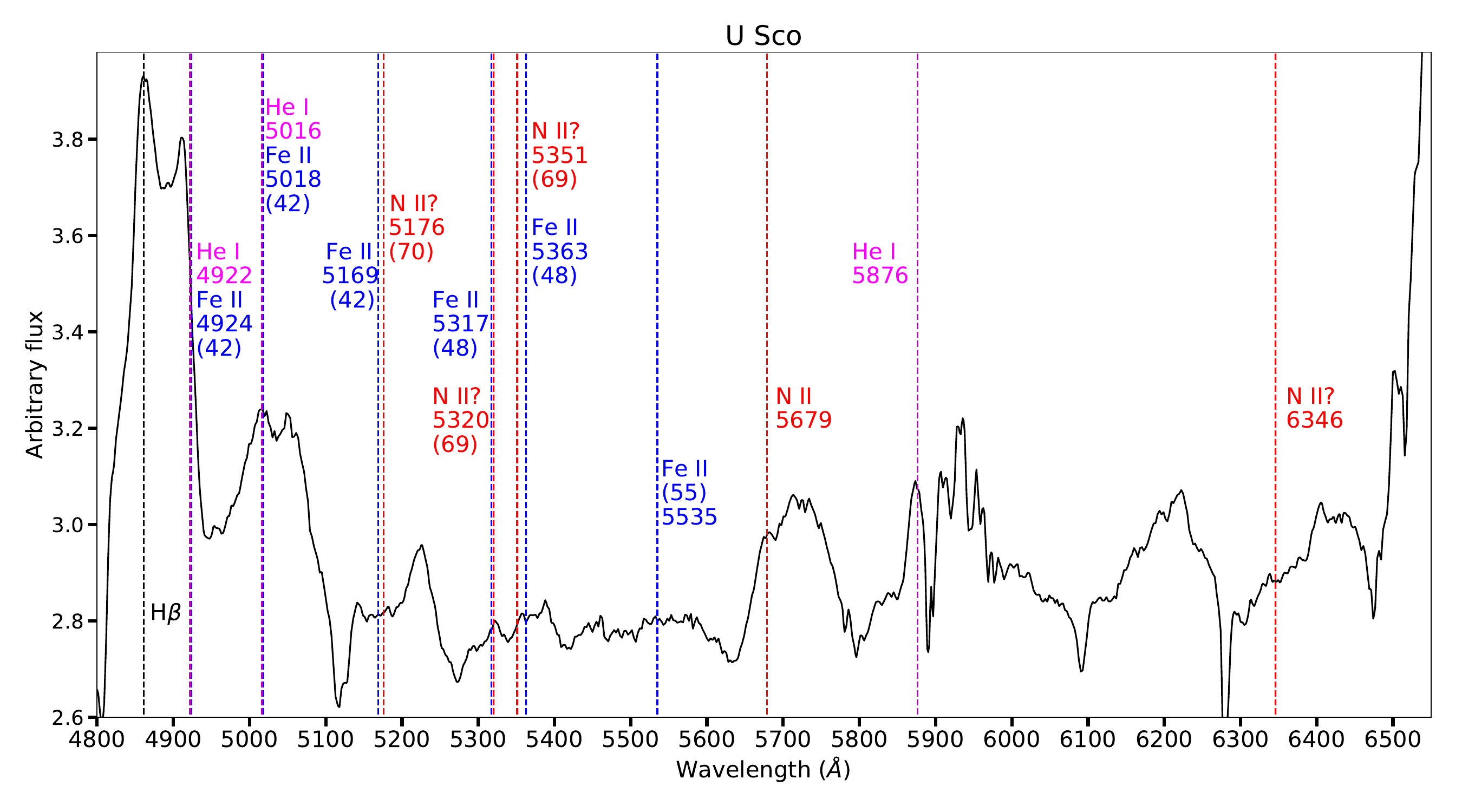}
\caption{A Zoom-in spectral plot of nova U Sco on day 0.7, focusing on the region of Fe II (42) and (48) multiplets, and other He I and N II lines.}
\label{Fig:U_Sco_spec_zoom}
\end{center}
\end{figure*}

\onecolumn
\begin{table}
\centering
\caption{Spectroscopic observation logs of nova T~Pyx (2011).}
\begin{tabular}{ccccc}
\hline
Source & date & $ t - t_0$ & Resolving power & $\lambda$ Range \\ 
 & &(days) &  & (\AA)\\
\hline
Astrourf & 2011-04-15.8 & 1.6 & 1000 & 3850-6850\\
Astrourf & 2011-04-16.8 & 2.6 & 1000 & 3850-9000\\
Astrourf & 2011-04-17.8 & 3.6 & 1000 & 3850-6850\\
Astrourf & 2011-04-18.8 & 4.6 & 1000 & 3850-6850\\
Astrourf & 2011-04-19.8 & 5.6 & 1000 & 3850-6850\\
Astrourf & 2011-04-21.8 & 7.6 & 600 & 4200-7500\\
Astrourf & 2011-04-29.8 & 15.6 & 1000 & 3850-6850\\		
SMARTS/CHIRON & 2011-05-02.2 & 18 & 27000 & 3800-9530\\
SMARTS/CHIRON & 2011-05-15.2 & 31 & 27000 & 3800-9530\\
SMARTS/CHIRON & 2011-06-01.2 & 48 & 27000 & 3800-9530\\
SMARTS/CHIRON & 2011-06-10.2 & 57 & 27000 & 3800-9530\\
SMARTS/CHIRON & 2011-06-25.2 & 73 & 27000 & 3800-9530\\
SMARTS/CHIRON & 2011-07-01.2 & 79 & 27000 & 3800-9530\\
SMARTS/CHIRON & 2011-09-24.2 & 164 & 27000 & 3800-9530\\
\hline
\end{tabular}
\label{table:spec_log_T_Pyx}
\end{table}

\begin{table}
\centering
\caption{Spectroscopic observation logs of nova V339~Del.}
\begin{tabular}{ccccc}
\hline
Source & date & $ t - t_0$ & Resolving power & $\lambda$ Range \\ 
 & &(days) &  & (\AA)\\
\hline
ARAS & 2013-08-14.8	& 0.5 & 10000 & 4200-7300\\
ARAS & 2013-08-15.8	& 1.5 & 10000 & 4200-7300\\
ARAS & 2013-08-16.8	& 2.5 & 10000 & 4200-7300\\
ARAS & 2013-08-18.8	& 4.5 & 10000 & 4200-7300\\
ARAS & 2013-08-19.8	& 5.5 & 10000 & 4200-7300\\
ARAS & 2013-08-21.8	& 8.5 & 10000 & 4200-7300\\
ARAS & 2013-08-26.8	& 13.5 & 10000 & 4200-7300\\
ARAS & 2013-09-01.8	& 19.5 & 10000 & 4200-7300\\
ARAS & 2013-09-09.8	& 27.5 & 10000 & 4200-7300\\
ARAS & 2013-09-15.8	& 33.5 & 10000 & 4200-7300\\
ARAS & 2013-09-21.8	& 39.5 & 10000 & 4200-7300\\
ARAS & 2013-09-30.8	& 48.5 & 10000 & 4200-7300\\
ARAS & 2013-10-07.8	& 55.5 & 10000 & 4200-7300\\
ARAS & 2013-12-01.8	& 110.5 & 10000 & 4200-7300\\
\hline
\end{tabular}
\label{table:spec_log_V339_Del}
\end{table}

\begin{table}
\centering
\caption{Spectroscopic observation logs of nova V659~Sct (ASASSN-19aad).}
\begin{tabular}{ccccc}
\hline
Source & date & $ t - t_0$ & Resolving power & $\lambda$ Range \\ 
 & &(days) &  & (\AA)\\
\hline
ARAS & 2019-10-30 & 1.0 & 1000 & 3850-7250\\
SOAR/Goodman & 2019-11-03 & 4 & 1100 & 3850-7800\\
SAOR/Goodman & 2019-11-04 & 5 & 1100 & 3850-7800\\
SOAR/Goodman & 2019-11-05 & 6 & 1100 & 3850-7800\\
ARAS & 2019-11-08 & 9 & 1000 & 3850-7250\\
ARAS & 2019-11-18 & 19 & 1000 & 3850-7250\\
SALT/HRS & 2020-06-21 & 236 & 14000 & 4000-8800\\
SALT/HRS & 2020-07-01 & 246 & 14000 & 4000-8800\\
SALT/HRS & 2020-07-07 & 252 & 14000 & 4000-8800\\
\hline
\end{tabular}
\label{table:spec_log_V659_Sct}
\end{table}

\begin{table}
\centering
\caption{Spectroscopic observation logs of nova V1405~Cas.}
\begin{tabular}{ccccc}
\hline
Source & date & $ t - t_0$ & Resolving power & $\lambda$ Range \\ 
 & &(days) &  & (\AA)\\
\hline
ARAS & 2021-03-19 & 1 & 1500 & 3900–7300\\
ARAS & 2021-03-27 & 9 & 14000 & 4000–7400\\
ARAS & 2021-04-06 & 19 & 14000 & 3900–7500\\
ARAS & 2021-04-19 & 32 & 1100 & 3900–7500\\
ARAS & 2021-05-20 & 63 & 1100 & 3800–7500\\
ARAS & 2021-06-23 & 97 & 1100 & 3800-7500\\
ARAS & 2021-07-11 & 115 & 1000 & 3800–7500\\
ARAS & 2021-07-30 & 134 & 1100 & 3800–7500\\
ARAS & 2021-08-16 & 151 & 1000 & 3900–7300\\
ARAS & 2021-08-26 & 161 & 14000 & 3800–7800\\
ARAS & 2021-09-10 & 176 & 14000 & 3800–8900\\
ARAS & 2021-09-27 & 193 & 1100 & 3900–7300\\
ARAS & 2021-10-04 & 200 & 1100 & 3800–8900\\
ARAS & 2021-10-16 & 212 & 1100 & 3800–8900\\
ARAS & 2021-11-09 & 236 & 1500 & 3900–7800\\
ARAS & 2021-12-22 & 279 & 1100 & 3800–7200\\
ARAS & 2022-02-19 & 338 & 1100 & 3800-7500\\
ARAS & 2022-03-14 & 361 & 1100 & 3900-7400\\
\hline
\end{tabular}
\label{table:spec_log_V1405_Cas}
\end{table}

\begin{table}
\centering
\caption{Spectroscopic observation logs of nova V606~Vul.}
\begin{tabular}{ccccc}
\hline
Source & date & $ t - t_0$ & Resolving power & $\lambda$ Range \\ 
 & &(days) &  & (\AA)\\
\hline
Asiago & 2021-07-17.12 & 1.8 & TBA & 3850-7750\\
Asiago & 2021-07-18.8 & 3.5 & TBA & 3800-7300\\			
Asiago & 2021-07-19.8 & 4.5 & TBA & 3850-7750\\
ARAS & 2021-07-21.5 & 6 & 1000 & 4000-7500\\
ARAS & 2021-07-24.5 & 9 & 1000 & 4000-7500\\
ARAS & 2021-07-26.5 & 11 & 1000 & 4000-7500\\
ARAS & 2021-07-27.5 & 12 & 1000 & 4000-7500\\
Asiago & 2021-07-28.8 & 13.5 & TBA & 3800-7300\\
ARAS & 2021-07-30.5 & 15 & 1000 & 4000-7500\\
SOAR & 2021-07-30.5 & 15 & 1000 & 800-7500\\
ARAS & 2021-08-06.7 & 22.4 & 1000 & 3800-7400\\
Asiago & 2021-08-18.8 & 33.5 & TBA & 3850-7750\\
ARAS & 2021-08-25.5 & 41 & 1000 & 4000-7500\\
ARAS & 2021-09-04.5 & 51 & 1000 & 4000-7500\\
ARAS & 2021-10-25.5 & 101 & 1000 & 4000-7500\\			
ARAS & 2021-11-01.5 & 109 & 1000 & 4000-7500\\
ARAS & 2021-11-06.5 & 114 & 1000 & 4000-7500\\
ARAS & 2021-11-10.5 & 118 & 1000 & 4000-7500\\
ARAS & 2021-11-12.5 & 120 & 1000 & 4000-7500\\
ARAS & 2022-03-25.5 & 253 & 1000 & 4000-7500\\
Asiago & 2022-11-07.8 & 480 & TBA & 3850-7750\\
\hline
\end{tabular}
\label{table:spec_log_V606_Vul}
\end{table}

\begin{table}
\centering
\caption{Spectroscopic observation logs of nova Gaia22alz.}
\begin{tabular}{ccccc}
\hline
Source & date & $ t - t_0$ & Resolving power & $\lambda$ Range \\ 
 & &(days) &  & (\AA)\\
\hline
SALT/RSS & 2022-03-10 & 45 & 1500 & 4200–7300\\
SALT/HRS & 2022-03-15 & 49 & 14000 & 4000–9000\\
SALT/HRS & 2022-04-03 & 68 & 14000 & 4000–9000\\
SOAR/Goodman & 2022-04-09 & 76 & 1100 & 3800–7800\\
SOAR/Goodman & 2022-04-14 & 80 & 1100 & 3800–7800\\
SOAR/Goodman & 2022-04-21 & 86 & 1100 & 3800–7800\\
ARAS & 2022-04-30 & 95 & 1000 & 4000–7800\\
SOAR/Goodman & 2022-05-01 & 96 & 1100 & 3800–7800\\
ARAS & 2022-05-02 & 97 & 1000 & 4000–7800\\
SALT/HRS & 2022-05-08 & 103 & 14000 & 4000–9000\\
SALT/HRS & 2022-05-11 & 106 & 14000 & 4000–9000\\
SOAR/Goodman & 2022-05-22 & 118 & 1100 & 3800–7800\\
SOAR/Goodman & 2022-05-24 & 120 & 1100 & 3800–7800\\
SOAR/Goodman & 2022-06-09 & 135 & 1100 & 3800–7800\\
Magellan/IMACS & 2022-07-21 & 178 & 1500 & 3800–6600\\
SOAR/Goodman & 2022-08-15 & 203 & 1100 & 3800–7800\\
SOAR/Goodman & 2022-08-24 & 212 & 1100 & 3800–7800\\
SOAR/Goodman & 2022-09-17 & 236 & 1100 & 3800–7800\\
SOAR/Goodman & 2022-10-08 & 257 & 1100 & 3800–7800\\
SOAR/Goodman & 2022-10-18 & 267 & 1100 & 3800–7800\\
SOAR/Goodman & 2022-10-29 & 278 & 1100 & 3800–7800\\
SOAR/Goodman & 2022-12-01 & 310 & 1100 & 3800–7800\\
SOAR/Goodman & 2023-01-20 & 360 & 1100 & 3800–7800\\
SOAR/Goodman & 2023-02-28 & 399 & 1100 & 3800–7800\\
SOAR/Goodman & 2023-03-15 & 414 & 1100 & 3800–7800\\
SOAR/Goodman & 2023-03-25 & 424 & 1100 & 3800–7800\\
SOAR/Goodman & 2023-04-04 & 434 & 1100 & 3800–7800\\
SOAR/Goodman & 2023-06-06 & 497 & 1100 & 3800–7800\\
\hline
\end{tabular}
\label{table:spec_log_Gaia22alz}
\end{table}

\begin{table}
\centering
\caption{Spectroscopic observation logs of nova V612~Sct.}
\begin{tabular}{ccccc}
\hline
Source & date & $ t - t_0$ & Resolving power & $\lambda$ Range \\ 
 & &(days) &  & (\AA)\\
\hline
Asiago & 2017-06-25 & 6 & TBA & 4000-7800\\
ARAS & 2017-06-29 & 10 & 1000 & 4400-7400\\
ARAS & 2017-06-30 & 11 & 1000 & 4400-7400\\
ARAS & 2017-07-02 & 12 & 1000 & 4400-7400\\
ARAS & 2017-07-06 & 16 & 1000 & 4400-7400\\
ARAS & 2017-07-07 & 17 & 1000 & 4400-7400\\
\hline
\end{tabular}
\label{table:spec_log_V612_Sct}
\end{table}

\begin{table}
\centering
\caption{Spectroscopic observation logs of nova FM~Cir.}
\begin{tabular}{ccccc}
\hline
Source & date & $ t - t_0$ & Resolving power & $\lambda$ Range \\ 
 & &(days) &  & (\AA)\\
\hline
SMARTS/CHIRON & 2018-01-20.5 & 2.8 & 27000 & 4100-8900\\
SMARTS/CHIRON & 2018-01-22.5 & 4.8 & 27000 & 4100-8900\\
SMARTS/CHIRON & 2018-01-23.5 & 5.8 & 27000 & 4100-8900\\
SMARTS/CHIRON & 2018-01-25.5 & 7.8 & 78000 & 4100-8900\\
SMARTS/CHIRON & 2018-01-31.5 & 13.8 & 78000 & 4100-8900\\
\hline
\end{tabular}
\label{table:spec_log_FM_Cir}
\end{table}

\begin{table}
\centering
\caption{Spectroscopic observation logs of nova V407~Lup.}
\begin{tabular}{ccccc}
\hline
Source & date & $ t - t_0$ & Resolving power & $\lambda$ Range \\ 
 & & (days) &  & (\AA)\\
\hline
ARAS & 2016-09-24.5 & 0.5 & 1000 & 4200-7300\\
VLT-Pucheros & 2016-09-29.5 & 5.5 & 20000 & 4250-6900\\
VLT-UVES & 2016-10-02.5 & 8.5 & 59000 & 3060-9460\\
ARAS & 2016-10-04.5 & 10.5 & 1000 & 4200-7300\\
VLT-UVES & 2016-10-12.5 & 18.5 & 59000 & 3060-9460\\
\hline
\end{tabular}
\label{table:spec_log_V407_Lup}
\end{table}

\begin{table}
\centering
\caption{Spectroscopic observation logs of nova U~Sco (2022).}
\begin{tabular}{ccccc}
\hline
Source & date & $ t - t_0$ & Resolving power & $\lambda$ Range \\ 
 & & (days) &  & (\AA)\\
\hline
SALT/HRS & 2022-06-07.0 & 0.3 & 14000 & 4000-8800\\
ARAS & 2022-06-07.2 & 0.5 & 1000 & 3800-7200\\
ARAS & 2022-06-07.4 & 0.7 & 1000 & 3800-7200\\
ARAS & 2022-06-07.8 & 1.1 & 1000 & 3800-7200\\
ARAS & 2022-06-08.3 & 1.6 & 1000 & 3800-7200\\
\hline
\end{tabular}
\label{table:spec_log_U_Sco}
\end{table}

\end{document}